\documentclass[aps,twocolumn,showpacs,showkeys,superscriptaddress]{revtex4}
\usepackage{epsfig}
\usepackage{amsmath,amssymb}
\usepackage{graphicx}

\newcommand{\bastar}{\begin{eqnarray*}}
\newcommand{\eastar}{\end{eqnarray*}}
\newskip\humongous \humongous=0pt plus 1000pt minus 1000pt

\newif\ifdtup

\relax
\newcommand{\bea}{\begin{eqnarray}}
\newcommand{\eea}{\end{eqnarray}}
\newcommand{\X}{{\vec X}}
\newcommand{\pro}{\partial}
\newcommand{\n}{\hat n}
\newcommand{\mn}{{\mu\nu}}
\newcommand{\oneg}{\displaystyle\frac{1}{g}}

\newcommand{\bq}{\bar q}
\newcommand{\bu}{\bar u}
\newcommand{\bd}{\bar d}
\newcommand{\bs}{\bar s}

\newcommand{\bg}{\bar g}

\newcommand{\bR}{\bar R}
\newcommand{\bB}{\bar B}
\newcommand{\bG}{\bar G}
\newcommand{\F}{\vec F}
\newcommand{\W}{\vec W}

\newcommand{\hF}{\hat F}

\newcommand{\A}{{\vec A}}
\newcommand{\hA}{{\hat A}}
\newcommand{\cA}{{\cal A}}
\newcommand{\cC}{{\cal C}}
\newcommand{\valpha}{{\vec \alpha}}

\newcommand{\hn}{{\hat n}}
\newcommand{\hD}{{\hat D}}

\newcommand{\nn}{\nonumber}

\begin{document}
\title{Abelian Decomposition and Glueball-Quarkonium Mixing in QCD}
\bigskip

\author{Pengming Zhang}
\affiliation{Institute of Modern Physics, Chinese Academy of
Science, Lanzhou 730000, China}
\author{Li-Ping Zou}
\affiliation{Institute of Modern Physics, Chinese Academy of
Science, Lanzhou 730000, China}
\author{Y. M. Cho}
\email{ymcho0416@gmail.com}
\affiliation{Institute of Modern Physics, Chinese Academy of
Science, Lanzhou 730000, China}
\affiliation{Center for Quantum Spacetime, Sogang University,
Seoul 04107, Korea} 
\affiliation{School of Physics and Astronomy,
Seoul National University, Seoul 08826, Korea}

\date{\today}

\begin{abstract}

The Abelian decomposition of QCD which decomposes
the gluons to the color neutral binding gluons (the neurons) 
and the colored valence gluons (the chromons) gauge 
independently naturally generalizes the quark model to 
the quark and chromon model which could play the central 
role in hadron spectroscopy. We discuss the color reflection 
symmetry, the fundamental symmetry of  the quark and 
chromon model, and explain how it describes the glueballs 
and the glueball-quarkonium mixing in QCD. We present 
the numerical analysis of glueball-quarkonium mixing 
in $0^{++}$, $2^{++}$, and $0^{-+}$ sectors below 
2 GeV, and show that in the $0^{++}$ sector $f_0(500)$ 
and $f_0(1500)$, in the $2^{++}$ sector $f_2(1950)$, 
and in the $0^{-+}$ sector $\eta(1405)$ and $\eta(1475)$ 
could be identified as predominantly the glueball states. 
We discuss the physical implications of our result.

\end{abstract}

\maketitle

\section{Introduction}

An important issues in hadron spectroscopy is 
the identification of the glueballs. The general wisdom 
is that QCD must have the glueballs made of gluons \cite{frit,pf,kogut}, and several models of glueball have 
been proposed \cite{jaff,roy,shif,coyne,chan,corn}. Moreover, the lattice QCD was able to construct 
the low-lying glueballs based on the first principles of 
QCD dynamics \cite{lqcd1,lqcd2}, and Particle Data Group (PDG) has accumulated a large number of hadronic states which do not seem to fit to the simple quark model as 
the glueball candidates \cite{pdg}. 

In spite of the huge efforts to identify the glueballs experimentally, however, so far the search for 
the glueballs has not been so successful \cite{prep1,prep2,prep3,math,ochs}. There are two 
reasons for this. First, theoretically there has been no consensus on how to construct the glueballs. This has 
made it difficult to predict what kind of glueballs we could expect. To see this consider the two leading models 
of glueballs, the bag model and the constituent gluon 
model.  

The bag model identifies the glueballs as the gauge 
invariant combinations of the gluon fields confined 
in a bag \cite{bag,jaff,roy}. In this model the confinement 
is imposed by the boundary condition of the bag, where 
the interaction among the confined gluons is described 
by the perturbative gluon exchange. On the other hand 
in the constituent gluon model the glueballs are identified 
as the color singlet bound states of the color octet ``constituent gluons", where the confinement is enforced 
by the confining potential \cite{coyne,chan}. 

Intuitively these models look reasonable and 
attractive, although they have their own advantages 
and disadvantages. They were able to show 
the existence of glueballs. But they have not been so 
successful to pinpoint exactly what are the glueball 
states and tell us how can we verify them. 

The other reason is that it is not clear how to identify 
the glueballs experimentally. This is partly because 
they could mix with quarkoniums, so that we must take 
care of the possible mixing to identify the glueballs 
experimentally \cite{prep1,prep2,prep3,math,ochs}. 
This is why we have very few candidates of the glueballs 
so far, compared to huge hadron spectrum made of 
quarks listed in PDG. 

This makes the search for the glueballs an urgent issue
in high energy physics, and we have detectors (e.g., 
GlueX at Jefferson Lab and PANDA at FAIR) specifically designed to search for the glueballs \cite{jlab,panda}. 
To have a successful identification of glueballs, however, 
we must have a better picture of the glueball. 

The Abelian decomposition of QCD allows us to do 
that \cite{prd80,prl81,prd81,duan}. It decomposes 
the QCD gauge potential to the Abelian restricted 
potential which has the full color gauge degrees of 
freedom and the gauge covariant valence potential which describes the colored gluons (the chromons) in a gauge independent way. Moreover, it decomposes the restricted potential further to the non-topological Maxwell part 
which describes the color neutral binding gluons 
(the neurons) and the topological Dirac part which 
describes the non-Abelian monopole. 

This tells that there are two types of gluons which play different roles. The neurons play the role of the binding gluons which bind the colored source, while the chromons play the role of the colored source of QCD. So we can 
view QCD as the restricted QCD (RCD) made of restricted potential which has the chromons as the colored source. 

We emphasize that this is against the common wisdom 
that all gluons (because of the gauge symmetry) are 
equal, carrying the same color charge. The Abelian decomposition tells that this is not true, and tells us 
how to separate the colored chromons from the color 
neutral neurons unambiguously.

Moreover, the Abelian decomposition allows us to study 
the role of the monopole, and prove that it is the monopole 
which is responsible for the confinement in lattice 
QCD \cite{kondo1,kondo2,cundy1,cundy2}. As importantly, 
it allows us to calculate the QCD effective action and 
demonstrate the monopole condensation gauge
independently \cite{prd01a,prd13,ijmpa14}. 

But what is most important for our purpose is that it allows 
us to have a clear picture of glueballs with which we can 
identify them. This is because the chromons play the role 
of the constituent gluons while the neurons bind them, 
after the confinement sets in. So we can construct 
the glueballs with a finite number of chromons as 
the constituent. This generalizes the quark model to 
the quark and chromon model which provides a new 
picture of hadrons \cite{prd80,prl81,prd15}. 

The quark model has been very successful. But the quark and chromon model has many advantages. It predicts 
new hadronic states, for example the hybrid hadrons 
made of quarks and chromons. More importantly, it 
provides a clear picture of glueballs and their mixing 
with the iso-singlet quarkoniums, and allows us to 
calculate the gluon content of the mixed states. 

Of course, the constituent gluon model can also do that, 
but this model can not tell the difference between 
the binding gluons and the constituent gluons. To understand this consider the hydrogen atom (or
any atom) in QED. Obviously we have photon as well 
as electron (and proton and neutron) in it, but only 
the electron determines the atomic structure of 
the atom in the periodic table. The photon plays no 
role in the atomic structure. It is there in the form of
the electromagnetic field to provide the binding, 
not as the constituent which determines the atomic structure of the atom. So we need not know how 
many of them are in the atom to determine it's place 
in the periodic table.

Exactly the same way the proton has quarks and 
gluons, but only the three quarks become 
the constituent. The gluons inside the proton do 
not play any role in the baryonic structure of 
the proton which determine the place of proton in 
the hadron spectroscopy. This means that they must 
be the ``binding" gluons, not the ``constituent" 
gluons, which (just like the photons in the hydrogen 
atom) provide only the binding of the quarks in 
the proton. If so, what are the constituent gluons, 
and how can we distinguish them from the binding 
gluons? Obviously the constituent gluon model does 
not provide the answer. 

The Abelian decomposition naturally resolves this 
difficulty. It tells that there are indeed two types of 
gluons, the neurons and the chromons, and in general 
only the chromons could be treated as the constituent gluons \cite{prd80,prl81}. This is because the neurons 
(like the photons) provide the binding force for colored objects, but the chromons (just like the quarks) 
become the colored source which make bound states 
in QCD.  So (with few exceptions) only the chromons 
could be qualified to be the constituent of hadrons.

In this picture the proton has no constituent chromons. 
But we emphasize that this does not mean proton does 
not contain the chromons at all. Clearly the three 
valence quarks which make up the proton can exchange chromons among themselves. Moreover, proton could 
have an infinite number of ``the sea chromons", just 
as they have the sea quarks. But obviously these 
chromons do not play the role of the constituent. 

In the quark and chromon model one could (in principle) 
construct an infinite number of glueballs with chromons. 
So one might ask why experimentally we have not so many
candidates of them. One could think of two reasons why 
this is so. First, the glueballs made of chromons have 
an intrinsic instability \cite{prd13,ijmpa14}. So they have 
broad widths, broader than the normal hadronic decay 
width. This means that they have a relatively short 
life-time. So only the low-lying glueballs could actually 
be observed experimentally. This is because the chromons, 
unlike the quarks, tend to annihilate each other in 
the chromo-electric background. This must be contrasted 
with quarks, which remain stable inside the hadrons. 

This is closely related to the asymptotic freedom 
(anti-screening) of gluons. It is well known that in QED 
the strong electric background tends to generate 
the pair creation of electrons, which makes the charge 
screening \cite{schw,prl01,prd01b}. But in QCD gluons 
and quarks play opposite roles in the asymptotic freedom. 
The quarks enhance the screening while the gluons 
diminish it to generate the anti-screening \cite{wil,pol}. 
In fact in the presence of a chromo-electric background 
the chromon loop generates a negative imaginary part 
but the quark loop generates a positive imaginary part in 
the QCD effective action. This tells that the chromo-electric 
field tends to generate the pair annihilation of the 
chromons \cite{sch,prd02,jhep04,prd13,ijmpa14}.

Second, in our model the glueballs inevitably mix with 
quarkoniums, so that in general they do not appear as 
mass eigenstates. So, to identify the glueballs, we have 
to consider the possible mixing with the quarkoniums. 
This makes the experimental identification of glueballs 
a nontrivial matter. This is another reason why 
the experimental identification of the glueballs so far 
has not been so successful.

Of course, in rare cases we could have the pure glueballs
called the oddballs \cite{coyne,prd15}. This is because
some of the chromoballs have the quantum number 
$J^{PC}$ which can not be made possible with $q\bq$.
In this case there is no $q\bq$ which could mix with
the oddballs, so that they may exist as pure chromoballs.
This makes the identification of the oddballs an important 
issue in QCD.  

In a recent paper we have discussed the general framework 
of hadron spectroscopy based on the quark and chromon 
model, and showed how the model can explain 
the glueball-quarkonium mixing and allow us to identify 
the glueballs \cite{prd15}. The present paper is the sequel
of this work in which we extend the preceding work and 
discuss the numerical analysis of the glueball-quarkonium 
mixing in more detail to help identify the glueballs without 
ambiguity. 

Our analysis makes it clear that the chromoballs play 
the central role in the meson spectroscopy, although 
in general they do not appear as mass eigenstates. In particular, our analysis tells that the chromoball-quarkonium mixing makes a deep influence on the $q\bq$ octet-singlet 
mixing. In fact in the quark and chromon model the $q\bq$ 
octet-singlet mixing can not be discussed without 
the chromoball-quarkonium mixing, because 
the chromoball-quarkonium mixing inevitably induces 
the octet-singlet mixing. 

The paper is organized as follows. In Section II we review 
the Abelian decomposition which decomposes the gluons 
to the color neutral neurons and the colored chromons 
to justify the quark and chromon model. In Section III 
we discuss the color reflection symmetry which replaces 
the non-Abelian gauge symmetry and becomes 
the fundamental symmetry of the quark and chromon 
model. In Section IV we explain how  the chromoballs, 
the bound states of chromons, can be understood as 
the glueballs in the quark and chromon model. In Section 
V we discuss the glueball-quarkonium mixing mechanism. 
In Section VI we present the numerical analysis of 
the low-lying glueball-quarkonium mixing in $0^{++}$, 
$2^{++}$, and $0^{-+}$ sectors below 2 GeV, and 
show that $f_0(1500)$, $f_2(1950)$, $\eta(1405)$, 
and $\eta(1475)$ become the strong candidates of 
glueballs. Finally in the last section we discuss 
the physical implications of our analysis.    

\section{Abelian Decomposition of Gluons: Neurons and Chromons} 

Before we discuss the Abelian decomposition we have 
to know why we need it. Consider the proton. The quark model tells that it is made of three quarks, but obviously 
we need the gluon to bind them. On the other hand 
the quark model tells that there is no ``valence" gluon 
inside the proton which can be a constituent of 
the proton. If so, what is the ``binding" gluon inside 
the proton, and how do we distinguish it from the valence gluon? 

Another motivation is the Abelian dominance, which 
asserts that the Abelian part of QCD is responsible for 
the color confinement \cite{thooft,prd00}. This must 
be true, because the non-Abelian (off-diagonal) part describes the colored gluons which are destined to be confined. Since the confined prisoner can not be 
the confining agent (the jailer), only the Abelian part 
can play the role of the confiner. But what is the Abelian part, and how do we separate it? 

The Abelian decomposition decomposes the QCD 
gauge potential to the restricted (Abelian) part and 
the valence (colored) part gauge independently. Consider the SU(2) QCD first, and let $(\hn_1,\hn_2,\hn_3=\hn)$ 
be an arbitrary local orthonormal basis. To make 
the Abelian decomposition we choose any direction, 
for example $\hn$, to be the Abelian direction and 
impose the isometry to project out the restricted 
potential $\hA_\mu$ \cite{prd80,prl81,prd81}
\bea
&D_\mu \hn=(\pro_\mu+g\A_\mu \times) \hn=0,  \nn\\
&\A_\mu \rightarrow \hA_\mu
=A_\mu \n-\oneg \n \times \pro_\mu \n
=\cA_\mu+\cC_\mu,  \nn\\
&\cA_\mu=A_\mu \n,~~\cC_\mu
=-\oneg \n \times \pro_\mu \n,
~~A_\mu=\hn \cdot \A_\mu. 
\label{ap}
\eea
The Abelian projection has the followings features. 
First, $\hA_\mu$ is precisely the potential which 
leaves the Abelian direction invariant under 
the parallel transport. Second, it is made of two parts, 
the non-topological (Maxwellian) $\cA_\mu$ which describes the color neutral gluon (the neuron) and 
the topological (Diracian) $\cC_\mu$ which describes 
the non-Abelian monopole \cite{prl80}. 
Third, the decomposition is gauge independent. We can 
rotate $\n$ to any direction and still get exactly the same 
decomposition. 

With this we have
\bea
& \hF_\mn = (F_\mn+ H_\mn)\hn, \nn\\
&F_\mn = \pro_\mu A_\nu-\pro_\nu A_\mu,
~H_\mn =-\dfrac1g \n \cdot (\pro_\mu \n \times \pro_\nu \n).
\label{rf} 
\eea 
This tells the followings. First, $\hF_\mn$ has only the Abelian 
component. Second, $\hF_\mn$ has a dual structure, made of
non-topological $F_\mn$ and topological $H_\mn$. 

With (\ref{ap}) we can recover the full QCD potential adding
the non-Abelian (colored) part $\X_\mu$ which describes 
the colored gluons (the chromons) \cite{prd80,prl81}
\bea
&\A_\mu = \hA_\mu + \X_\mu,  
~~~~\hn \cdot \X_\mu=0.
\label{adec}
\eea
Under the infinitesimal gauge transformation 
\bea
\delta \A_\mu=\oneg  D_\mu \vec \alpha, 
~~~\delta \hn_i= -\vec \alpha \times \hn_i, 
\eea 
we have 
\bea 
&\delta \hA_\mu=\dfrac1g \hD_\mu \vec \alpha, 
~~~\delta \X_\mu = - \valpha \times \X_\mu,
\eea 
where $\hD_\mu=\pro_\mu+g\hA_\mu\times$. 
This tells that $\hA_\mu$ has the full SU(2) gauge 
degrees of freedom, even though it is restricted. 
Moreover, $\vec X_\mu$ becomes gauge covariant. 

Notice that, although the neuron is given by the Abelian 
component of $\A_\mu$, the chromon is not given by 
the non-Abelian component of $\A_\mu$. This is because 
the Abelian decomposition decomposes $\A_\mu$ to 
the neuron, chromon, and the topological monopole.
So the topological part plays an essential role in 
the Abelian decomposition.

With the restricted potential we can construct the restricted 
QCD (RCD) which has the full non-Abelian gauge symmetry 
but is simpler than the QCD 
\bea 
&{\cal L}_{RCD} =-\dfrac{1}{4} \hF^2_\mn
=-\dfrac{1}{4} F_\mn^2 \nn\\
&+\dfrac1{2g} F_\mn \hn \cdot (\pro_\mu \hn \times \pro_\nu \hn)
-\dfrac1{4g^2} (\pro_\mu \hn \times \pro_\nu \hn)^2,
\label{rcd}
\eea
which describes the Abelian subdynamics of QCD. Since RCD contains the non-Abelian monopole degrees 
explicitly, it provides an ideal platform for us to study 
the monopole dynamics gauge independently.

From (\ref{adec}) we have
\bea
\vec{F}_{\mu\nu}&=&\hat F_{\mu \nu} + \hD _\mu \X_\nu 
- \hD_\nu \X_\mu + g\X_\mu \times \X_\nu. 
\eea 
With this we can express QCD by
\bea 
&{\cal L}_{QCD} = -\dfrac{1}{4} \F^2_{\mu\nu }
=-\dfrac{1}{4}\hF_\mn^2-\dfrac{1}{4}(\hD_\mu\X_\nu
-\hD_\nu\X_\mu)^2 \nn\\
&-\dfrac{g}{2} {\hat F}_{\mu\nu} \cdot (\X_\mu \times \X_\nu)
-\dfrac{g^2}{4} (\X_\mu \times \X_\nu)^2. 
\label{2ecd} 
\eea
This is the extended SU(2) QCD (ECD) which confirms that 
QCD can be viewed as RCD made of the binding gluons, 
which has the chromons as its source \cite{prd80,prl81}. 

The Abelian decomposition is more complicated but straightforward. Since SU(3) has rank two, it has two 
Abelian directions. Let $\n_i~(i=1,2,...,8)$ be an arbitrary local orthonormal SU(3) basis, and choose $\n_3=\n$ 
and $\n_8=\n'$ to be the Abelian directions. Make 
the Abelian projection by
\bea
D_\mu \n=0.
\label {3cp}
\eea
This automatically guarantees \cite{prl80}
\bea
D_\mu \n'=0,~~~\n'=\dfrac1{\sqrt 3} \n*\n.
\eea
where $*$ denotes the $d$-product. This is because 
SU(3) has two vector products, the anti-symmetric 
$f$-product and the symmetric $d$-product.

Solving (\ref{3cp}), we have the Abelian projection which 
projects out the binding potential,
\bea
&\A_\mu \rightarrow \hA_\mu=A_\mu \hn+A_\mu' \hn'
-\oneg \hn\times \pro_\mu \hn
-\oneg \hn'\times \pro_\mu \hn' \nn\\
&=\sum_p \dfrac23 \hA_\mu^p,~~~(p=1,2,3),    \nn\\
&\hA_\mu^p=A_\mu^p \n^p
-\oneg \n^p \times \pro_\mu \n^p
=\cA_\mu^p+\cC_\mu^p, \nn\\
&\cA_\mu^p=A_\mu^p \n^p,
~~~\cC_\mu^p=-\oneg \n^p \times \pro_\mu \n^p,  \nn\\
&A_\mu^1=A_\mu,~~~A_\mu^2=-\dfrac{1}{2}A_\mu
+\dfrac{\sqrt 3}{2}A_\mu',  \nn\\
&A_\mu^3=-\dfrac{1}{2}A_\mu-\dfrac{\sqrt 3}{2}A_\mu',  
~~~\n^1=\n,   \nn\\
&\n^2=-\dfrac{1}{2} \n +\dfrac{\sqrt 3}{2} \n',
~~~\n^3=-\dfrac{1}{2} \n -\dfrac{\sqrt 3}{2} \n',
\label{cp3}
\eea
where the sum is the sum of the Abelian 
directions of three SU(2) subgroups made of 
$(\n_1,\n_2,\n^1),~(\n_6,\n_7,\n^2),~(\n_4,-\n_5,\n^3)$.
Notice that the three $\hA_\mu^p$ are not mutually
independent.

Under the infinitesimal gauge transformation 
\bea
&\delta \A_\mu=\oneg  D_\mu \vec \alpha,
~\delta \hn= -\vec \alpha \times \hn, 
~\delta \hn'= -\vec \alpha \times \hn',
\eea 
we have \cite{prd80,prl81,prd81}
\bea 
&\delta \hA_\mu=\dfrac1g \hD_\mu \vec \alpha.
\eea
This confirms that $\hA_\mu$ has the full SU(3) gauge 
degrees of freedom. 

From this we have the restricted field strength made of 
the binding potential
\bea
&\hF_\mn=\dfrac23 \sum_p \hF_\mn^p
=\dfrac23 \sum_p (F_\mn^p+H_\mn^p)^2,   \nn\\
&F_\mn^p=\pro_\mu A_\nu^p-\pro_\nu A_\mu^p,  \nn\\
&H_\mn^p=-\dfrac1g \n^p \cdot (\pro_\mu \n^p 
\times \pro_\nu \n^p), 
\eea
and obtain the restricted QCD (RCD) which has the full SU(3) 
gauge symmetry \cite{prd80,prl81,prd81}
\bea
&{\cal L}_{RCD} =-\dfrac{1}{4} \hF_\mn^2  
= -\dfrac{1}{6} \sum_p (F_\mn^p+H_\mn^p)^2.
\label{rcd3}
\eea
Just like the SU(2) RCD it has a dual structure, made of 
$F_\mn^p$ and $H_\mn^p$. 

\begin{figure}
\includegraphics[height=2.5cm, width=6cm]{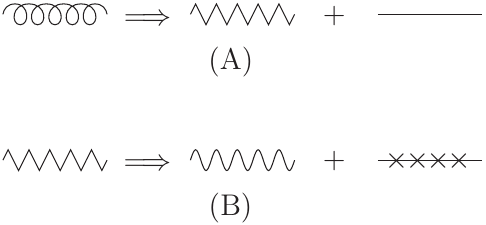}
\caption{\label{cdec} The Abelian decomposition of 
the gluons. The gauge potential is decomposed to 
the restricted potential (kinked line) and the chromon (straight line) in (A), and the restricted potential is 
further decomposed to the neuron (wiggly line) and 
the monopole (spiked line) in (B).}
\end{figure}

Adding the valence part $\X_\mu$ which describes 
the chromons to the binding potential we have 
the Abelian decomposition of the SU(3) gauge 
potential \cite{lambda}
\bea
&\A_\mu=\hat A_\mu+\X_\mu
=\sum_p (\dfrac23 \hA_\mu^p+\W_\mu^p), \nn\\
&\X_\mu= \sum_p \W_\mu^p,  \nn\\
&\W_\mu^1= X_\mu^1 \n_1+ X_\mu^2 \n_2,
~~~\W_\mu^2=X_\mu^6 \n_6 + X_\mu^7 \n_7,  \nn\\
&\W_\mu^3= X_\mu^4 \n_4 +X_\mu^5 \n_5.
\label{cdec3}
\eea
Again, under the gauge transformation we 
have \cite{prd80,prl81,prd81}
\bea
&\delta \hA_\mu = \oneg  \hD_\mu \vec \alpha,
~~~\delta \X_\mu = - \valpha \times \X_\mu.
\label{cgt}
\eea
This confirms that $\vec X_\mu$ becomes gauge covariant.  
Moreover, this tells that the chromons $\X_\mu$  can be 
decomposed to the three SU(2) chromons $\W_\mu^p$. 
But unlike $\hA_\mu^p$, they are mutually independent. 
So we have two neurons and six chromons in SU(3) QCD. 
The Abelian decomposition has also been known as 
the Cho decomposition, Cho-Duan-Ge decomposition, 
or Cho-Faddeev-Niemi decomposition \cite{fadd,shab,gies,zucc,kondor}. 

From (\ref{cdec3})  we have
\bea
&\hD _\mu \X_\nu=\sum_p \hD_\mu^p \W_\nu^p,
~~~\hD_\mu^p=\pro_\mu+ g \hA_\mu^p \times,   \nn\\
&\X_\mu\times \X_\nu
=\sum_{p,q} \W_\mu^p \times \W_\nu^q,  \nn\\
&\vec{F}_{\mu\nu}=\hF_\mn + \hD _\mu \X_\nu 
-\hD_\nu \X_\mu + g\X_\mu \times \X_\nu  \nn\\
&=\sum_p \big[\dfrac23 \hF_\mn^p
+ (\hD_\mu^p \W_\nu^p-\hD_\mu^p \W_\nu^p) \big]  \nn\\
&+\sum_{p,q}\W_\mu^p \times \W_\nu^q,
\eea
so that we can express the SU(3) QCD as \cite{prd01a,prd13,ijmpa14}
\bea
&{\cal L}_{ECD}={\cal L}_{RCD}
-\dfrac{1}{4}(\hD_\mu\X_\nu-\hD_\nu\X_\mu)^2 \nn\\
&-\dfrac{g}{2} (\hD_\mu \X_\nu
-\hD_\nu \X_\mu) \cdot (\X_\mu \times \X_\nu)\nn\\
&-\dfrac{g}{2} {\hF}_\mn \cdot (\X_\mu \times \X_\nu)
-\dfrac{g^2}{4} (\X_\mu \times \X_\nu)^2 \nn\\
&= \sum_p \Big[-\dfrac{1}{6} (\hF_\mn^p)^2
-\dfrac{1}{4} (\hD_\mu^p \W_\nu^p
- \hD_\nu^p \W_\mu^p)^2 \nn\\
&-\dfrac{g}{2} \hF_\mn^p \cdot (\W_\mu^p 
\times \W_\nu^p) \Big]   
-\sum_{p,q} \dfrac{g^2}{4} (\W_\mu^p 
\times \W_\mu^q)^2 \nn\\
&-\sum_{p,q,r} \dfrac{g}2 (\hD_\mu^p \W_\nu^p
- \hD_\nu^p \W_\mu^p)
\cdot (\W_\mu^q \times \W_\mu^r)  \nn\\
&-\sum_{p\ne q} \dfrac{g^2}{4} \Big[(\W_\mu^p 
\times \W_\nu^q) \cdot (\W_\mu^q \times \W_\nu^p)  \nn\\
&+(\W_\mu^p \times \W_\nu^p)
\cdot (\W_\mu^q \times \W_\nu^q) \Big].
\label{3ecd}
\eea
This is the SU(3) ECD, which is mathematically identical 
to QCD. Adding an extra term or subtracting any existing term is strictly forbidden.  

\begin{figure}
\includegraphics[height=4.5cm, width=7cm]{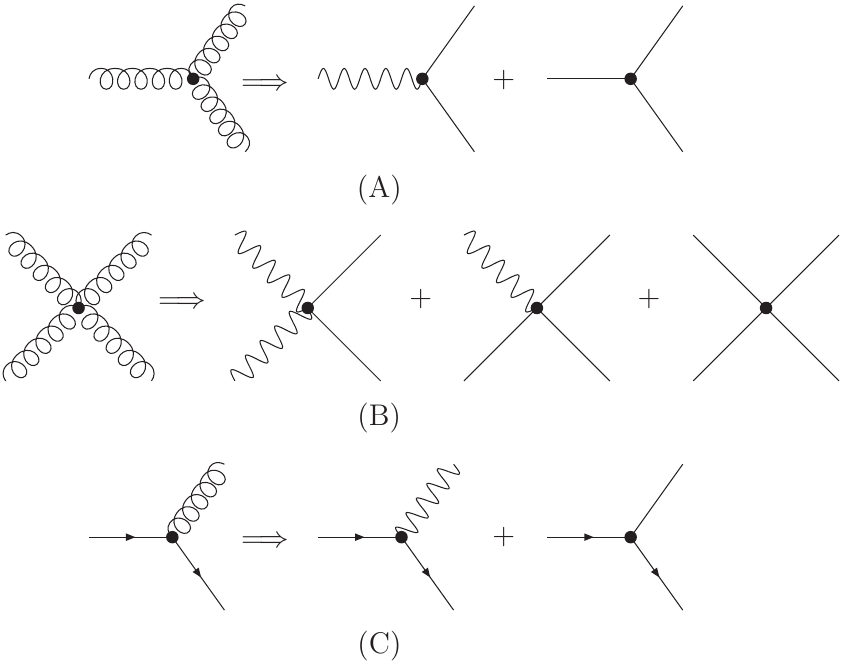}
\caption{\label{ecdint} The Abelian decomposition of Feynman diagrams in SU(3) QCD. The three-point and 
four-point gluon vertices are decomposed in (A) and (B), and the quark-gluon vertices are decomposed in (C). 
Notice that the monopole does not appear in the Feynman diagram, since it does not represent a dynamical degree.}
\end{figure}

We can easily add  quarks in the Abelian decomposition,
\bea
&{\cal L}_{q} =\sum_k\bar \Psi_k 
(i\gamma^\mu D_\mu-m) \Psi_k \nn\\
&= \sum_k \Big[\bar \Psi_k (i\gamma^\mu \hD_\mu-m) \Psi_k
+\dfrac{g}{2} \X_\mu
\cdot \bar \Psi_k (\gamma^\mu \vec t) \Psi_k \Big] \nn\\
&=\sum_{p,k} \Big[\bar \Psi_k^p 
(i\gamma^\mu \hD_\mu^p-m) \Psi_k^p
+\dfrac{g}{2} \W_\mu^p \cdot \bar \Psi_k^p
(\gamma^\mu \vec \tau^p) \Psi_k^p \Big], \nn\\
&\hD_\mu = \pro_\mu + \dfrac{g}{2i} {\vec t}\cdot \hA_\mu,
~~\hD_\mu^p=\pro_\mu
+\dfrac{g}{2i} {\vec \tau^p}\cdot \hA_\mu^p,
\label{qlag}
\eea
where $m$ is the mass, $k$ is the flavour index,  
$\vec t$ is the color generators of the quark triplet corresponding to the chromons $\X_\mu$, $p$ denotes 
the color generators of the quarks corresponding to 
three SU(2) subgroups of SU(3), and $\Psi_k^p$ 
represents the three SU(2) quark doublets ({i.e., 
(r,b), (b,g), and (g,r) doublets) of the (r,b,g) quark 
triplet. 

The Abelian decomposition is summarized graphically. 
This is shown in Fig. \ref{cdec}, where the gluons are decomposed to the restricted potential and the chromon potential in (A), and the restricted potential is 
decomposed further to the non-topological neuron 
potential $\cA_\mu$ and the topological monopole 
potential $\cC_\mu$ in (B). 

Although the Abelian decomposition does not change 
QCD, it reveals important hidden structures of QCD. 
First of all, it tells that there are two types of gluon, 
the neuron and chromon, which play totally different role. This means that there should be two types of gluon jets, 
the neuron jet and the chromon jet, which in principle 
could be tested and confirmed by experiment. Without 
the Abelian decomposition we could not tell this because 
all gluons are treated on equal footing. 

Second, it allows us to decompose the QCD Feynman diagram in such a way that the conservation of color is 
made explicit. This is shown in Fig. \ref{ecdint}. In (A) 
the three-point gluon vertex is decomposed to two 
vertices made of one neuron plus two chromons and 
three chromons. In (B) the four-point gluon vertex is decomposed to three vertices made of one neuron 
plus three chromons, two neurons plus two chromons, 
and four chromons. In (C) the quark-gluon vertex is decomposed to the quark-neuron vertex and quark-chromon vertex. 

Notice that three-point vertex made of three neurons 
or two neurons and one chromon, and four-point vertex made of three or four neurons are forbidden by 
the conservation of color. Moreover, the quark-neuron interaction does not change the quark color, but 
the quark-chromon interaction changes the quark color.

Another point is that the monopole does not appear 
in the diagram for the following reasons. First, after 
the confinement (the monopole condensation) sets in, 
the monopole part disappears completely. So, in 
the perturbative regime (inside the hadrons where 
the asymptotic freedom applies) only the neurons and chromons contribute to the Feynman diagrams. But 
the more fundamental reason is that the monopole, 
as the topological degree of QCD, does not become 
a dynamical (i.e., propagating) degree. So it can not 
appear in the Feynman diagram \cite{prd01a}. 

\begin{figure}
\includegraphics[height=4.5cm, width=8cm]{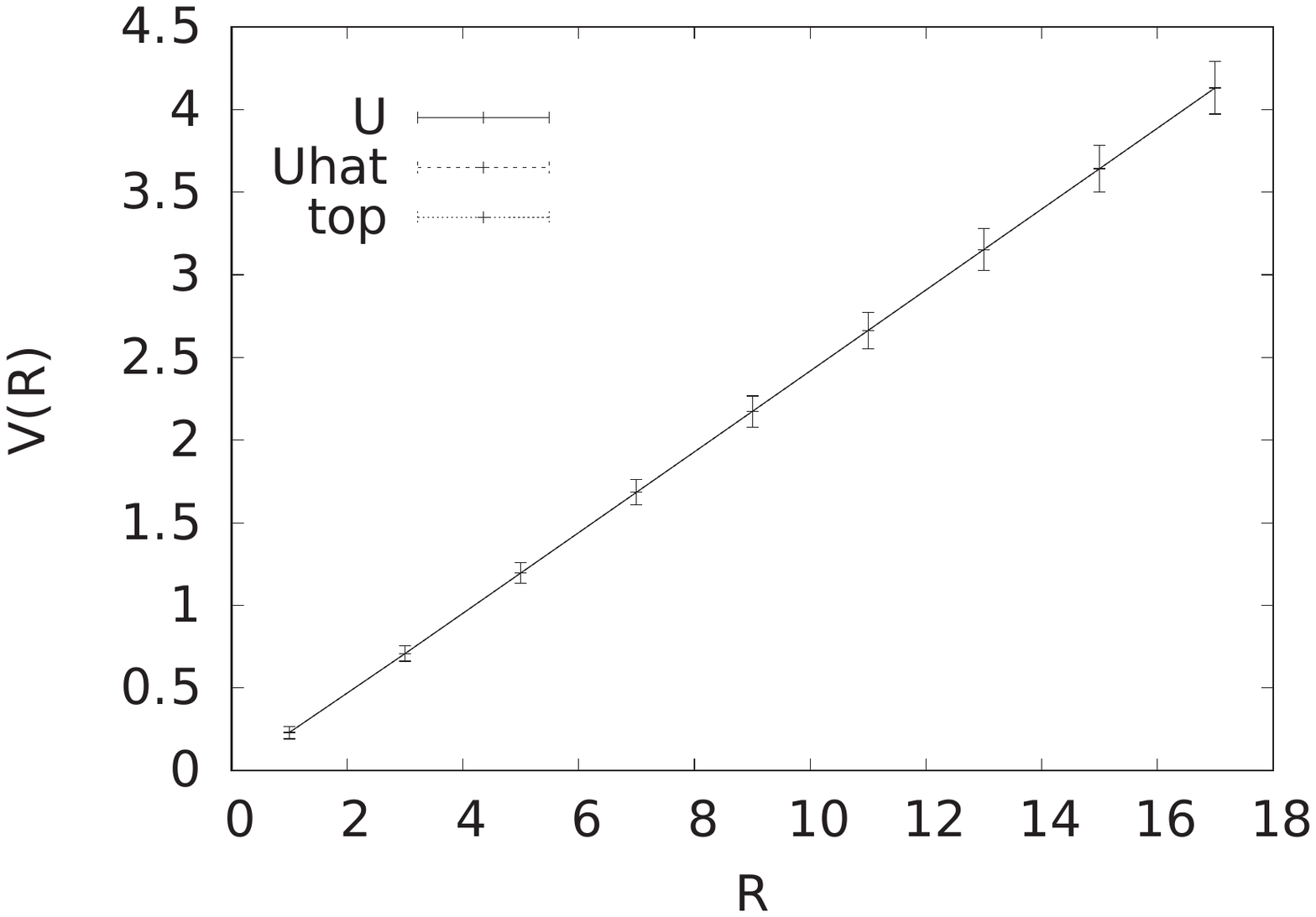}
\caption{\label{cundy} The lattice QCD calculation which 
establishes the monopole dominance in Wilson loop. Here 
the solid, dotted, and dashed lines are obtained with 
the full potential, the restricted potential, and 
the monopole potential, respectively.}
\end{figure}

Third, the Abelian decomposition of SU(3) QCD reveals 
the Weyl symmetry of the SU(3) QCD, and shows that 
the theory is invariant under the permutation of three 
SU(2) subgroups, or equivalently three colors of SU(3). 
Indeed (\ref{cp3}), (\ref{3ecd}), and (\ref{qlag}) clearly 
show that they are invariant under the permutation of 
three SU(2) subgroups. The Weyl group of SU(N) is 
the $N!$-elements permutation group of $N$ colors. In general the Abelian decomposition allows us to express 
the SU(N) QCD explicitly in the Weyl symmetric form. 
This is very important, because this allows us to express 
the SU(N) QCD effective action in terms of the SU(2) 
QCD effective action.   
  
In the non-perturbative regime, the Abelian decomposition 
allows us to demonstrate the monopole dominance, that 
it is the monopole which confines the color. In fact implementing the Abelian decomposition on lattice, we 
can calculate the contribution of the Wilson loop with 
the full potential, the restricted potential, and 
the monopole potential separately, and show that 
the monopole potential produces the confining 
force \cite{kondo1,kondo2,cundy1,cundy2}. The recent lattice result obtained with the Abelian decomposition 
is copied in Fig. \ref{cundy}, which shows that all three potentials produce exactly the same confining force\cite{cundy1,cundy2}. Clearly this proves that the neuron and chromon do not contributes to the Wilson loop 
integral.  

The lattice result demonstrates the monopole dominance, 
that the monopole is essential for the confinement. However, it does not show how the monopole confines 
the color. To show this we have to calculate the effective action of QCD. The Abelian decomposition and 
the resulting ECD provides us an ideal platform for us 
to calculate the QCD effective action gauge 
independently \cite{prd13,ijmpa14}. 

This is because field theoretically the Abelian 
decomposition puts QCD in the background field 
formalism \cite{prd01a,dewitt,pesk}. So we can treat 
the restricted potential and the valence potential 
as the slow varying classical field and the fluctuating quantum field, and calculate the QCD effective action 
in the presence of the monopole background imposing 
the gauge invariance. 

This allows us to show that the true QCD vacuum is 
given by the monopole condensation, more precisely 
the monopole-antimonopole pair condensation \cite{prd13,ijmpa14}. The SU(3) QCD effective 
potential is shown in Fig. \ref{su3ep}. This strongly 
implies that it is the monopole condensation which 
generates the mass gap and the confinement in QCD. 

\begin{figure}
\begin{center}
\includegraphics[height=5cm, width=7cm]{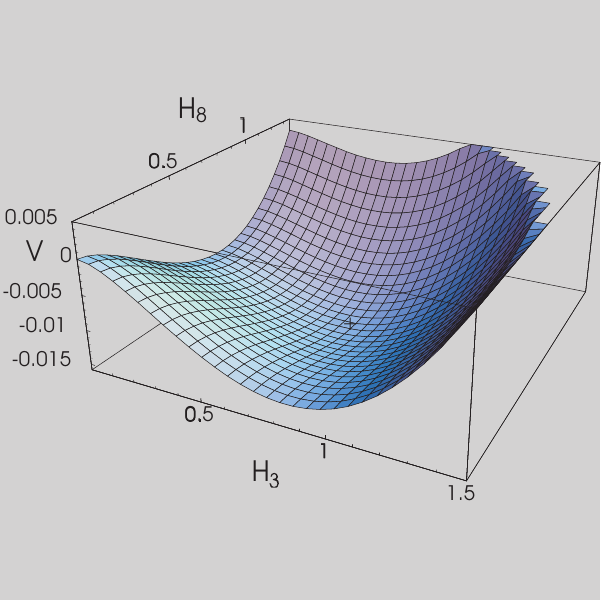}
\caption{\label{su3ep} The one-loop effective potential 
of SU(3) QCD which demonstrates the monopole condensation. The potential is obtained by integrating 
out the chromons  in the presence of a constant 
monopole background.}  
\end{center}
\end{figure}

The most important point of the Abelian decomposition 
for our purpose in this paper, however, is that it 
generalizes the quark model to the quark and chromon model \cite{prd15}. This is because the chromons, being colored, are naturally qualified to become the constituent 
of hadrons. In contrast the neurons, being neutral, play 
the role of photons which provides the binding in QED. 

This leads us to the quark and chromon model where 
the colored quarks and chromons become the constituent 
of hadrons. This gives us a new picture of hadron spectroscopy. Moreover, this provides us a clear picture 
of the glueballs and their mixing with quarkoniums, and helps us to identify the glueballs experimentally.  

To understand how the quark and chromon model 
works, it is important to understand that the Abelian decomposition reduces the complicated non-Abelian 
color gauge symmetry to the discrete symmetry made 
of finite elements called the color reflection symmetry 
which becomes the fundamental symmetry of the quark 
and chromon model \cite{prd80,prl81}. This is very important, because this color reflection symmetry 
plays the role of the gauge symmetry but is much 
easier to handle. So we discuss the color reflection symmetry in detail in the next section.     

\section{Color Reflection Invariance and Weyl Symmetry of ECD}

As we have emphasized, the Abelian decomposition 
is gauge independent. On the other hand, the selection 
of the Abelian direction amounts to the gauge fixing 
which breaks the gauge symmetry. But this does not 
break the gauge symmetry completely, because we 
have a residual discrete symmetry called the color 
reflection symmetry even after the Abelian 
decomposition \cite{prd80,prl81,prd13}.

The importance of this residual symmetry comes from 
the following observation. First, this plays the role of 
the gauge symmetry after the Abelian decomposition. 
Second, this symmetry is much simpler than the color 
gauge symmetry. This tells that the Abelian decomposition
reduces the complicated non-Abelian gauge symmetry 
to a simple discrete symmetry which is much easier to 
handle. So we discuss the color reflection symmetry first. 

Consider the SU(2) QCD first and make the color 
reflection, the $\pi$-rotation of the SU(2) basis along 
the $\n_2$-direction which inverts the color direction 
$\n$,  
\bea
(\n_1,\n_2,\n) \rightarrow (-\n_1,\n_2,-\n).
\label{2cref}
\eea 
Obviously this is a gauge transformation which should 
not change the physics. On the other hand, under 
the color reflection (\ref{2cref}) we have \cite{prd13}
\begin{gather}
\hA_\mu \rightarrow \hA_\mu^{(c)}
= -A_\mu \n-\oneg \n \times \pro_\mu \n,  \nn\\
A_\mu \rightarrow A_\mu^{(c)}= -\n \cdot \A_\mu=-A_\mu.  
\label{2ncrt}
\end{gather}
Moreover, we have
\begin{gather}
\X_\mu \rightarrow \X_\mu^{(c)}
=-(X_\mu^1~\n_1-X_\mu^2~\n_2),  \nn
\end{gather}
or, in the complex notation 
\begin{gather}
R_\mu=\frac1{\sqrt 2}(X_\mu+iX_\mu^2) \nn\\
\rightarrow R_\mu^{(c)}=-\bR_\mu
=-\frac1{\sqrt 2}(X_\mu-iX_\mu^2),
\label{2ccrt}
\end{gather}
where $R_\mu$ denotes the red chromon. 

But since the isometry condition (\ref{ap}) is insensitive 
to (\ref{2cref}), we have two different Abelian decompositions imposing the same isometry, 
\begin{gather}
\A_\mu= \hA_\mu+\X_\mu,
~~~~\A_\mu= \hA_\mu^{(c)}+\X_\mu^{(c)}, 
\end{gather}
without changing the physics. This is why the color 
reflection (\ref{2cref}) becomes a discrete symmetry of 
QCD after the Abelian decomposition \cite{prd80,prl81}. 

To understand the meaning of this, notice that 
the neuron potential $A_\mu$ change the signature, 
while the topological part remains invariant. Moreover 
the chromon changes to the complex conjugate partner (together with the change of the signature), which 
changes the chromon to anti-chromon and flips 
the sign of the chromon charge.  

This is what is expected. In the absence of the topological part (\ref{2ecd}) describes QED which is coupled to 
the massless charged vector field where the neuron plays the role of the photon. And in QED it is well known that 
the photon has negative charge conjugation quantum number. So it is natural that $A_\mu$ in SU(2) QCD 
changes the signature under the color reflection. 
Similarly we can argue that $A_\mu$ changes the signature 
under the parity \cite{prd13}.   

As importantly, (\ref{2ccrt}) tells that the physics should 
not change when we change the chromon to anti-chromon,
because they are the color reflection partner. This means 
that they can not be separately discussed in QCD and 
should always play exactly the same amount of role. This 
is the reason why the color should become unphysical 
and confined, which makes QCD totally different from 
QCD \cite{prd80,prl81,prd13}. 

In the fundamental representation the color reflection 
(\ref{2cref}) is given by the 4 element subgroup of SU(2)
made of \cite{prd80,prl81}
\bea
&C_1=\left(\begin{array}{cc}
1 & 0 \\ 0 & 1 \\ \end{array} \right),
~~C_2=\left(\begin{array}{cc}
-1 & 0 \\ 0 & -1 \\ \end{array} \right), \nn\\
&C_3=\left(\begin{array}{cc}
0 & 1 \\ -1 & 0 \\ \end{array} \right),
~~C_4=\left(\begin{array}{cc}
0 & -1 \\ 1 & 0 \\ \end{array} \right).
\label{crg2}
\eea
This can be expressed by
\begin{gather}
C_k=D_a R_b, ~~(a=1,2;~b=1,2;~k=1,2,...,4),  \nn\\
D_1=\left(\begin{array}{cc}
1 & 0 \\ 0 & 1 \\ \end{array} \right),
~~~D_2=\left(\begin{array}{cc}
-1 & 0 \\ 0 & -1 \\ \end{array} \right)  \nn\\
R_1=\left(\begin{array}{cc}
1 & 0 \\ 0 & 1 \\ \end{array} \right),
~~~~R_2=\left(\begin{array}{cc}
0 & 1 \\ -1 & 0 \\ \end{array} \right),
\end{gather}
which contains the diagonal subgroup made of $D_1$ 
and $D_2$. And this becomes the residual symmetry 
of the SU(2) quark doublet $(r,b)$ after the Abelian decomposition. Notice that $R_2$ plays the role of 
the generator of the color reflection group.

As for the gluons which form the adjoint representation 
the color reflection can be simplified further for two reasons. 
First, the diagonal subgroup has no effect on the adjoint 
representation. Second, the color reflection changes $\n$ 
to $-\n$ and $(\n_1,\n_2)$ to $(-\n_1,\n_2)$. So, the gluon 
triplet is decomposed to two independent representations.

Indeed, for the neuron we have 
\begin{gather}
R_2:~~A_\mu \rightarrow -A_\mu. 
\label{2nrep}
\end{gather}
But for the chromon we have
\begin{gather}
R_2:~~(\X_\mu,\X_\mu^{(c)}) 
\rightarrow -(\X_\mu^{(c)},\X_\mu), \nn
\end{gather}
or equivalently
\begin{gather}
R_2:~~(R_\mu,\bR_\mu) \rightarrow -(\bR_\mu,R_\mu).
\label{2crep}
\end{gather}
This confirms that the neuron and chromon transform 
independently, forming a one-dimensional and 
two-dimensional representations under the color 
reflection. This drastically simplifies the non-Abelian 
gauge symmetry. 

For SU(3) the fundamental representation the color reflection group is made of 24 elements subgroup 
of SU(3) given by \cite{prd80,prl81,kondor}
\begin{gather}
C_k=D_a R_b,  \nn\\
(a=1,2,3,4;~b=1,2,...,6;~k=1,2,...,24),   \nn\\
D_1=\left(\begin{array}{ccc}
1 & 0 & 0 \\ 0 & 1 & 0 \\ 0 & 0 & 1 \\
\end{array} \right),
~~~D_2=\left(\begin{array}{ccc}
-1 & 0  & 0 \\ 0 & -1 & 0 \\ 0 & 0 & 1 \\
\end{array} \right),  \nn\\
D_3=\left(\begin{array}{ccc}
1 & 0  & 0 \\ 0 & -1 & 0 \\ 0 & 0 & -1 \\
\end{array} \right), 
D_4=\left(\begin{array}{ccc}
-1 & 0 & 0 \\ 0  & 1 & 0 \\ 0 & 0 & -1 \\
\end{array} \right),  \nn\\
R_1=\left(\begin{array}{ccc}
1 & 0 & 0 \\ 0 & 1 & 0 \\ 0 & 0 & 1 \\
\end{array} \right),
~~~~R_2=\left(\begin{array}{ccc}
0  & 1 & 0 \\ -1 & 0 & 0 \\ 0  & 0 & 1 \\
\end{array} \right), \nn\\
R_3=\left(\begin{array}{ccc}
1 & 0 & 0 \\ 0 & 0 & 1 \\ 0 & -1 & 0 \\
\end{array} \right),
~~~R_4=\left(\begin{array}{ccc}
0 & 0 & 1 \\ 0 & -1 & 0 \\ 1 & 0 & 0 \\
\end{array} \right), \nn\\
R_5=\left(\begin{array}{ccc}
0 & 1 & 0 \\ 0 & 0 & 1 \\ 1 & 0 & 0 \\
\end{array} \right),
~R_6=\left(\begin{array}{ccc}
0  & 0 & 1 \\ -1 & 0 & 0 \\ 0 & -1 & 0 \\
\end{array} \right),
\label{crg3}
\end{gather}
where the four $D$-matrices form the diagonal subgroup. 
This describes the residual symmetry of the quark triplet 
$(r,b,g)$ after the Abelian decomposition. Notice that 
here $R_2$ and $R_3$ play the role of the generator. 
For example, we have $R_5=R_3\cdot R_2$, 
$R_6=R_2\cdot R_3$, and $R_4=R_2\cdot R_3\cdot R_2$.

For the gluon octet which form the adjoint representation 
of SU(3) the color reflection can be simplified further. 
Just as in SU(2) QCD, the neurons and chromons 
transform separately, among themselves. To see exactly how they transform notice that the two neurons 
transform as
\begin{gather}
R_2:~~~~\left(\begin{array}{c} A_\mu \\ A_\mu' \end{array} \right)
\rightarrow \left(\begin{array}{cc}
-1 & 0 \\ 0 & 1 \\ \end{array} \right)  
\left(\begin{array}{c} A_\mu \\ A_\mu' \end{array} \right), \nn\\
R_3:\left(\begin{array}{c} A_\mu \\ A_\mu' \end{array} \right)
\rightarrow \left(\begin{array}{cc}
1/2 & \sqrt 3/2 \\ \sqrt 3/2 & -1/2 \\ 
\end{array} \right)
\left(\begin{array}{c} A_\mu \\ A_\mu' \end{array} \right),
\end{gather}
On the other hand, according to (\ref{cp3}) and 
(\ref{cdec3}) the two neurons form a (mutually 
dependent) triplet $(A_\mu^1,A_\mu^2,A_\mu^3)$. 
So in terms of the triplet the color reflection acts 
as follows,
\begin{gather}
R_2:~~(A_\mu^1,A_\mu^2,A_\mu^3) 
\rightarrow -(A_\mu^1,A_\mu^3,A_\mu^2),  \nn\\
R_3:~~(A_\mu^1,A_\mu^2,A_\mu^3) 
\rightarrow -(A_\mu^3,A_\mu^2,A_\mu^1),  \nn\\
R_4:~~(A_\mu^1,A_\mu^2,A_\mu^3) 
\rightarrow -(A_\mu^2,A_\mu^1,A_\mu^3),  \nn\\
R_5:~~(A_\mu^1,A_\mu^2,A_\mu^3) 
\rightarrow ~~(A_\mu^3,A_\mu^1,A_\mu^2),  \nn\\
R_6:~~(A_\mu^1,A_\mu^2,A_\mu^3) 
\rightarrow ~~(A_\mu^2,A_\mu^3,A_\mu^1).
\label{3ncrt}
\end{gather}
This tells that basically $R_2,~R_3,~R_4$ describe 
the permutations of two SU(2) neurons (up to the signature 
change), but $R_5,~R_6$ describe the cyclic permutations 
of three SU(2) neurons.
 
For the six chromons which form a sextet 
$(\W_\mu^1,\W_\mu^2,\W_\mu^3,\W_\mu^{1(c)},\W_\mu^{2(c)},\W_\mu^{3(c)})$ we can express them as 
three (red, blue, and green) colored chromons of 
the SU(2) subgroups by 
$(R_\mu,B_\mu,G_\mu,\bR_\mu,\bB_\mu,\bG_\mu)$. 
For these the color reflection acts as follows,
\begin{gather}
R_2:~~(R_\mu,B_\mu,G_\mu,\bR_\mu,\bB_\mu,\bG_\mu)  \nn\\
~~~~~~~~\longrightarrow (\bR_\mu,\bG_\mu,\bB_\mu,R_\mu,G_\mu,B_\mu),  \nn\\
R_3:~~(R_\mu,B_\mu,G_\mu,\bR_\mu,\bB_\mu,\bG_\mu)  \nn\\
~~~~~~~~\longrightarrow -(\bG_\mu,\bB_\mu,\bR_\mu,G_\mu,B_\mu,R_\mu),  \nn\\
R_4:~~(R_\mu,B_\mu,G_\mu,\bR_\mu,\bB_\mu,\bG_\mu)  \nn\\
~~~~~~~~\longrightarrow -(\bB_\mu,\bR_\mu,\bG_\mu,B_\mu,R_\mu,G_\mu),     \nn\\
R_5:~~(R_\mu,B_\mu,G_\mu,\bR_\mu,\bB_\mu,\bG_\mu)  \nn\\
~~~~~~~~\longrightarrow -(G_\mu,R_\mu,B_\mu,\bG_\mu,\bR_\mu,\bB_\mu),  \nn\\
R_6:~~(R_\mu,B_\mu,G_\mu,\bR_\mu,\bB_\mu,\bG_\mu)  \nn\\
~~~~~~~~\longrightarrow -(B_\mu,G_\mu,R_\mu,\bB_\mu,\bG_\mu,\bR_\mu).
\label{3ccrt}
\end{gather} 
Here $R_2,~R_3,~R_4$ describe the anti-chromon transformation (complex conjugation) plus permutations 
of two chromons, but $R_5,~R_6$ describe the cyclic permutations of three chromons (up to the signature change). 

The above discussion reveals another important difference between the neuron and chromon. Clearly (\ref{3ncrt}) tells 
that the neurons just permute and change the signature of 
the wave function, but (\ref{3ccrt}) tells that the chromons 
change to anti-chromons, under the color reflection. In other words, just like the photon in QED there is no anti-neurons 
in QCD. In contrast, the chromons have the anti-chromon 
partners. This is because the neurons are neutral but 
the chromons are colored, so that the neuron wave functions have real form, while the chromon wave 
functions have complex expression.    

At this point one might wonder if there is any relation between the color reflection group and Weyl group. 
For SU(3), the Weyl group is the six elements permutation group of three colors which has a three-dimensional representation given by
\begin{gather}
W_1=\left(\begin{array}{ccc}
1 & 0 & 0 \\ 0 & 1 & 0 \\ 0 & 0 & 1 \\
\end{array} \right),
~~~W_2=\left(\begin{array}{ccc}
0  & 1 & 0 \\ 1 & 0 & 0 \\ 0  & 0 & 1 \\
\end{array} \right), \nn\\
W_3=\left(\begin{array}{ccc}
1 & 0 & 0 \\ 0 & 0 & 1 \\ 0 & 1 & 0 \\
\end{array} \right),
~~~W_4=\left(\begin{array}{ccc}
0 & 0 & 1 \\ 0 & 1 & 0 \\ 1 & 0 & 0 \\
\end{array} \right), \nn\\
W_5=\left(\begin{array}{ccc}
0 & 1 & 0 \\ 0 & 0 & 1 \\1 & 0 & 0 \\
\end{array} \right),  
~~W_6=\left(\begin{array}{ccc}
0  & 0 & 1 \\ 1 & 0 & 0 \\ 0 & 1 & 0 \\
\end{array} \right), 
\label{wg3}
\end{gather}
which contains the cyclic $Z_3$ made of $W_1,~W_5,$ 
and $W_6$.

\begin{figure}
\includegraphics[height=4.5cm, width=8cm]{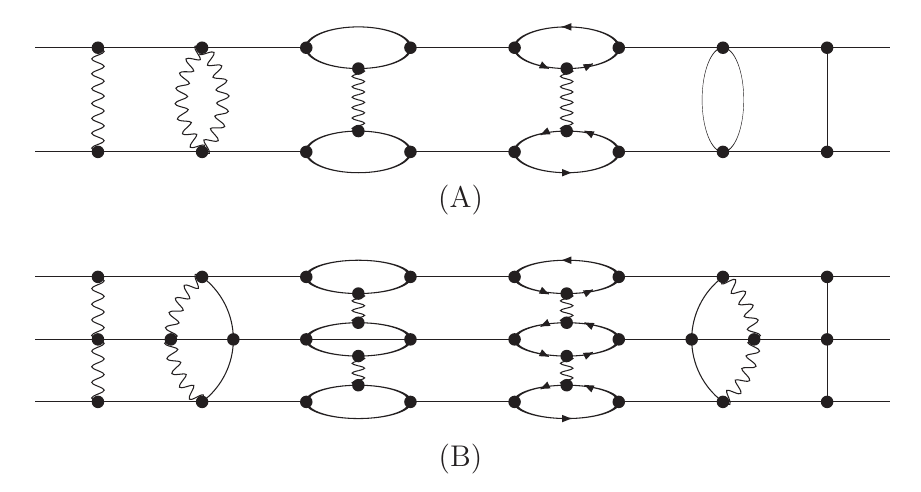}
\caption{\label{cball} The possible Feynman diagrams 
which bind the chromons. Two chromon binding is shown 
in (A), three chromon binding is shown in (B). The quarks 
are represented by the arrows.}
\end{figure}

This tells that the two groups are different. They have 
different origin. The Weyl group comes as the symmetry 
of the Abelian decomposition, but the color reflection group 
is the residual symmetry of the Abelian decomposition. 
Unlike the color reflection group (\ref{crg3}), the Weyl 
group (\ref{wg3}) is not a subgroup of SU(3). Moreover, 
the Weyl group has no complex conjugation operation 
which transforms the chromons to anti-chromons. On 
the other hand they have a common subgroup $Z_3$, 
the cyclic permutation group of three colors.  

Obviously both the color reflection group and the Weyl 
group should play a fundamental role in hadron spectroscopy. 
Only the color reflection invariant and Weyl invariant 
combinations of quarks and gluons can become physical 
in the quark and chromon model \cite{prd15}. On the other 
hand, the color reflection group plays a more fundamental 
role in the sense that it has the complex conjugation operation 
which transforms the chromons to anti-chromons.  

\section{Glueballs in Quark and Chromon Model: Chromoballs}

So far we have discussed the theoretical aspects of QCD 
which are exact. From now on we discuss their applications
which inevitably contains approximations, in particular 
the quark and chromon model in more detail. We first provide 
more argument  for the quark and chromon model. In 
the constituent gluon model the gauge invariant combinations 
of the octet gluons, $g\bg$ or $ggg$, have been thought 
to form the glueballs. This is because all gluons are treated 
equally in this model. 

As we have pointed out this has a critical defect. Obviously 
an important role of gluons is to provide the binging force of 
the colored objects. So, if all gluons become the constituent, 
it is very difficult to explain how they provide the binding. 
In the quark and chromon model, however, only the chromons become the constituent gluon. This is because only the chromons carry the color charge. In comparison the neurons, being color neutral, naturally assume the role of the binding gluons. 

\begin{figure}
\begin{center}
\includegraphics[height=4.5cm, width=8cm]{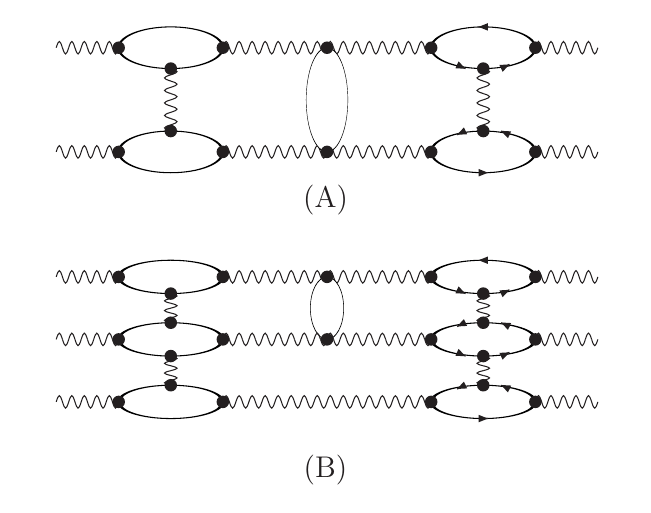}
\caption{\label{nball} The possible Feynman diagrams of 
the neuron interaction. Two neuron binding is shown in 
(A), three neuron binding is shown in (B). The quarks are 
represented by the arrows.}
\end{center}
\end{figure}

To clarify this point we compare the possible Feynman 
diagrams of two and three chromon interactions shown 
in Fig. \ref{cball} with the similar Feynman diagrams of 
neuron interactions shown in Fig. \ref{nball}. Clearly 
Fig. \ref{cball} looks very similar to the Feynman diagrams 
of $q\bq$ and $qqq$ bound states in the quark model. 
This means that the chromons, just like the quarks, can 
become the constituents of hadrons. In particular, this 
means that they could form chromoball bound states 
among themselves. 

On the other hand  Fig. \ref{nball} looks totally different 
from Fig. \ref{cball}. Obviously Fig. \ref{nball} looks very 
much like the photon self-interaction in QED. This is
because the neurons are not colored, so that they 
can interact only through the chromon or quark loops. 
So they actually play the role of ``the photons" in QCD
whose binding is much weaker than the chromon binding. 
This strongly support the quark and chromon model.

Of course, although the photons in QED do not form a bound 
state, there is still a possibility that the neuron binding is 
strong enough to form a bound state in QCD. Nevertheless 
it is natural to assume that, if the neurons form a bound state at all, they should form a very weakly bound state which would look like a bound state of two quarkoniums 
or a molecular state made of two light mesons. This means that there could be only a few neuroballs, the bound states made of neurons, maybe one or two at most. For this 
reason we will assume that only the chromons become 
the constituent in this paper.  

This, however, does not mean that they can not contribute to the binding. Clearly Fig. \ref{cball} and Fig. \ref{nball} show that both neurons and chromons can exchange chromons to make the binding. But we emphasize that 
there is a clear difference between the role of the chromon to be the constituent gluon and the exchange gluon. 

Now we discuss the characteristic features and new predictions of the quark and chromon model, and show 
how we can test the model experimentally. As we have explained, the most important change in this model is 
the replacement of the non-Abelian gauge group by 
the color reflection group. Indeed the color reflection symmetry becomes the backbone of the quark and 
chromon model \cite{prl81,prd81,prd15}. 

This simplifies the non-Abelian gauge invariance to the color 
reflection invariance. So only the color reflection invariant 
combinations of the chromons become gauge invariant and 
thus form the glueballs. This is why the color reflection group becomes so important in this model. 

To amplify this point we emphasize that the model reclassifies all hadrons in the quark model. For example, 
in the quark model mesons and baryons are viewed as 
color singlets made of $(3\times \bar 3)$ and 
$(3\times 3 \times 3)$ SU(3) quark triplets. But in 
the new model the quark triplets should be interpreted 
as the triplet of the color reflection group, not 
the color SU(3). So, in this reclassification the mesons 
and baryons acquire a different interpretation. This, of course, would not change the hadron spectrum much. 
But it clearly shows that the quark and chromon model sheds a new light on the old quark model, even for 
the hadrons made of quarks. 

Notice that for the meson classification the full color reflection symmetry becomes important, because 
the anti-chromons become an important ingredient. 
But for the baryons the Weyl symmetry plays the main 
role because baryons have only quarks, not anti-quarks.     

But obviously a best place where the quark and chromon 
model makes a big difference is the glueball and hybrid 
hadron. The model asserts that only the chromoball, 
the bound state of chromons, can become the glueball. 
In other words, the $gg$ and $ggg$ glueballs are actually the color singlets made of $(6\times \bar 6)$ and 
$(6\times 6\times 6)$ sextet chromons of the color reflection group, not $(8\times \bar 8)$ and 
$(8\times 8\times 8)$ SU(3) octets. This is the difference 
between the quark and chromon model and the constituent 
gluon model \cite{coyne,chan}. So in this model the quark 
triplet and chromon sextet of the color reflection group 
become the essential ingredients of hadrons. 

As importantly the model provides conceptually a clear picture of chromoball mixing with the quarkonium. 
Moreover, the model predicts the existence of the hybrid hadrons made of the quark triplets and chromon sextets. 
So studying the chromoballs and their mixing with quarkoniums and the hybrid hadrons predicted 
by the model we can test the quark and chromon model.        

In the quark and chromon model one expects infinite tower 
chromoballs, but experimentally we have not so many 
candidates of glueballs. There could be two possible 
explanations for this. First, they could easily mix with 
quakoniums, unless the conservation of quantum number 
forbids the mixing. This means that in reality the physical 
glueballs are mixed states, not pure chromoball states. 
Certainly this makes the experimental identification of 
the glueballs a non-trivial matter \cite{prd15}. 
 
Second, the chromoballs may have an intrinsic instability 
and decay faster than ordinary hadrons, which could 
make the experimental identification difficult. As we 
have pointed out, the chromons tend to annihilate 
each other in the color background, which has 
to do with the anti-screening of the color charge \cite{prd13,ijmpa14,prd15,sch,prd02,jhep04}. 

We can estimate the glueball partial decay width coming 
from this instability. According to the QCD one-loop 
effective action the chromon annihilation probability 
per unit volume per unit time is given 
by \cite{prd13,ijmpa14,prd15}
\bea
\Gamma_A=\sum_p \dfrac{11g^2}{96\pi} {\bar E}_p^2 
\times \dfrac{4\pi}{3\Lambda_{QCD}^3},
\label{gg}
\eea
where the sum is on three SU(2) subgroups and 
${\bar E}_p$ is the average chromo-electric field of 
each subgroup inside the glueballs. Now, if we 
choose $\alpha_s \simeq 0.4$, 
$\Lambda_{QCD}\simeq 339~{\rm MeV}$ (for three quark flavors), and $\bar E_p \simeq (g/\pi)\Lambda^2_{QCD}$ 
we have $\Gamma_A \simeq 398~{\rm MeV}$ \cite{pdg}. 
But notice that with 
$\Lambda_{QCD} \simeq 200~{\rm MeV}$, we have 
$\Gamma_A\simeq 235~{\rm MeV}$ \cite{pesk}.

Of course this is a rough estimate which depends on 
many things. For example, the $g\bg$ glueballs and 
$ggg$ glueballs may have different color field strengths 
and different sizes, and thus may have different life-time. But we emphasize that the above estimate is the partial decay width we expect from the asymptotic freedom, in addition to the ``normal" hadronic decay width. This strongly implies that in general the glueballs (in particular excited ones) are expected to have very short lifetime. 

This instability has another important implication. It has 
been widely believed that ``the gluon condensation" 
plays important role in QCD dynamics \cite{shif}. 
However, the gluon pair annihilation shown in (\ref{gg}) strongly suggests that this gluon condensation should become unstable, and thus can not last. Moreover, QCD already has the monopole condensation. This makes 
the gluon condensation highly improbable. 

\section{Glueball-Quarkonium Mixing}

In the preceding paper we have outlined how the chromoballs can mix with quarkoniums in the quark 
and chromon model, to show the viability of the above theoretical discussions \cite{prd15}. In this paper we 
discuss the mixing in more detail. The possible Feynman diagrams for the mixing are shown in Fig. \ref{mixing}, 
which tells that the mixing takes place not just between chromoballs and quarkoniums but also between 
the $cc$ and $ccc$ chromoballs, directly or through 
the virtual states made of neurons and/or molecular 
bound states of mesons. So in the mixing diagram 
the role of neuron and chromon is blurred.

Obviously the mixing influences the $q\bar{q}$ 
octet-singlet mixing in the quark model. So we review 
the  octet-singlet mixing in the quark model first. Let 
\bea
&\langle u\bar{u}|H|u\bar{u} \rangle_{Ex}
=\langle d\bar{d}|H|d\bar{d} \rangle_{Ex}=E, \nn\\
&\langle s\bar{s}|H|s\bar{s} \rangle_{Ex}=E'=E+\Delta, \nn\\
&\langle q'\bar{q'}|H|q\bar{q} \rangle_{An}=A,
~~~(\rm{for~all~q,q'}).
\eea
Now with
\bea
&|8 \rangle=\dfrac{|u\bar u\rangle+|d\bar d \rangle
-2|s\bar s\rangle}{\sqrt 6}, \nn\\
&|1 \rangle=\dfrac{|u\bar u \rangle+|d\bar d \rangle
+|s\bar s \rangle}{\sqrt 3},
\eea
we may obtain the following mass matrix for the $q\bar{q}$
which describes the octet-singlet mixing \cite{prd15},
\bea
&M^2=\left( \begin{array}{cc}  \langle 8 |H|8  \rangle
& \langle 8 |H| 1  \rangle  \\
\langle 1 |H| 8  \rangle   &   \langle 1 |H| 1  \rangle
\end{array} \right) \nn\\
&=\left( \begin{array}{cc}
E+\dfrac{2}{3}\Delta & -\dfrac{\sqrt{2}}{3}\Delta \\
-\dfrac{\sqrt{2}}{3}\Delta  & E+\dfrac{1}{3}\Delta +3A
\end{array} \right).
\label{osm}
\eea
Notice that $\Delta$-term is responsible for the mixing.
But we emphasize that this mixing among the quarks can 
not provide the correct octet-singlet mixing because the  
glueballs inevitably influence the quark octet-singlet 
mixing. This is evident in Fig. \ref{mixing}. 

\begin{figure}
\includegraphics[height=4.5cm, width=8cm]{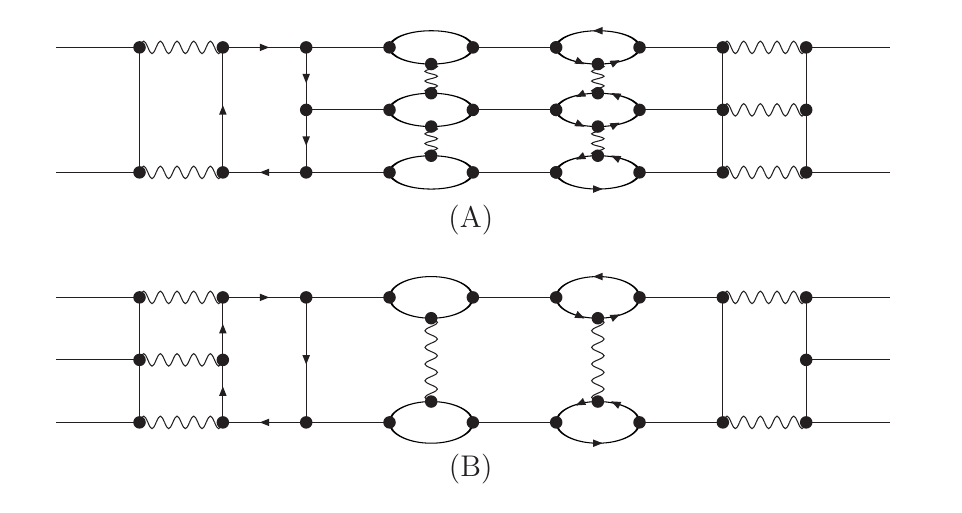}
\caption{\label{mixing} The possible glueball-quarkonium 
mixing diagrams. The $gg$ and $ggg$ chromoball mixing 
with quarkoniums are shown in (A) and (B).}
\end{figure}

Clearly we can generalize (\ref{osm}) to the following $3\times 3$ mixing matrix of one lightest chromoball 
$|G\rangle$ with the quark nonet \cite{prd15}  
\bea
M^2  =\left(\begin{array}{ccc}
E+\dfrac{2}{3}\Delta  & -\dfrac{\sqrt{2}}{3}\Delta  & 0 \\
-\dfrac{\sqrt{2}}{3}\Delta  & 
E+\dfrac{1}{3}\Delta +3A & \nu \\
0  &  \nu & G \end{array} \right).
\label{mm1}
\eea
Of course similar mixing matrix has been used in 
the constituent gluon model. As we have pointed out, however, this model has critical defects. And our quark 
and chromon model can be viewed as a new model 
which justifies the above mixing without such defects. 

In principle we should be able to calculate the parameters
in the mass matrix theoretically. For example, we could calculate $G$ using the gauge invariant current operator, 
or calculate the mixing parameter $\nu$ using the Feynman diagrams in our model. But in this paper we will fix 
the parameters with experimental data to see how well 
the mixing matrix can explain the glueball-quarkonium mixing.

Now, diagonalizing the mass matrix we can transform 
the unphysical states $(|8\rangle,|1\rangle,|G\rangle)$ 
to the mass eigenstates 
$(|m_1\rangle,|m_2\rangle,|m_3\rangle)$, and obtain 
the information on the chromon and quark contents of 
the physical states. Notice that, assuming that after 
the confinement the chromons acquire the constituent 
mass $\mu$, we can put $G=4 \mu^2$ (supposing 
the chromoball mass before the mixing is given by 
$\sqrt G= 2\mu$).     

We can easily generalize the mass matrix to 
the $4\times 4$ mixing
\bea
M^2 =\left( \begin{array}{cccc}
E+\dfrac{2}{3}\Delta  & -\dfrac{\sqrt{2}}{3} \Delta & 0 & 0 \\
-\dfrac{\sqrt{2}}{3}\Delta  & E+\dfrac{1}{3}\Delta 
+3A & \nu & \nu'\\
0  &  \nu & G & \epsilon  \\
0 & \nu' & \epsilon & G'  \end{array} \right),
\label{mm2}
\eea
to include one more chromoball state $|G' \rangle$. 
This has eight parameters, but we may reduce 
the parameters to seven by diagonalizing the $2\times 2$ chromoball mass matrix first and putting $\epsilon=0$. 
With this we can express $G$ and $G'$ by the chromon mass $\mu$ and put $G=4 \mu^2$ and 
$G'=4 \mu^2+\delta$ for two chromon bound state or 
$G'=9 \mu^2+\delta$ for three chromon bound state. 
Clearly (\ref{mm1}) and (\ref{mm2}) demonstrate that 
the chromoball-quarkonium mixing inevitably influences 
the quarkonium octet-singlet mixing. So we can not 
discuss the $q\bq$ spectroscopy without the chromoballs.
 
Diagonalizing the mass matrix we can figure out the quark 
and chromon contents of the mass eigenstates. Moreover, 
knowing the chromon content of the physical states, we can 
calculate the relative branching ratios of iso-singlet mesons 
made of heavy quarks, e. g., the $J/\psi$ radiative decay to 
the physical states in each channel. This is because these 
decays are the Okubo-Zweig-Iizuka (OZI) suppressed process which can only be made possible through 
the intermediate chromoball states. 

Let $\alpha_i$ be the parameters of the mixing matrix 
which determine the gluon content of physical states 
$|m_i\rangle$. We can predict the relative branching 
ratios of $J/\psi$ to $\gamma X$ decays among 
the physical states with $\alpha_i$, because these decays are induced by gluons.  So, for the S wave decay (i.e., 
for $0^{++}$ and $2^{++}$) we have \cite{prd15}
\bea
&R\Big(\dfrac{J/\psi \rightarrow \gamma X_k}{J/\psi\rightarrow 
\gamma X_i}\Big)=\Big(\dfrac{\alpha_k}{\alpha_i}\Big)^2
\Big(\dfrac{m^2_\psi-m^2_k}{m^2_\psi-m^2_i} \Big)^3,
\eea
but for the P wave decay (i.e., for $0^{-+}$) we expect 
to have
\bea
&R\Big(\dfrac{J/\psi \rightarrow \gamma X_k}{J/\psi\rightarrow 
\gamma X_i}\Big)=\Big(\dfrac{\alpha_k}{\alpha_i}\Big)^2
\Big(\dfrac{m^2_\psi-m^2_k}{m^2_\psi-m^2_i} \Big)^5,
\eea
where the last term is the kinematic phase space factor. 
Clearly this argument can also be applied to similar OZI 
suppressed decays of heavy $t \bar t$ or $b \bar b$ 
iso-singlet mesons. 

It must be pointed out that, although the chromoballs in 
general mix with the quarkoniums, in particular cases 
the pure chromoballs could exist. This is because 
some of the $gg$ chromoballs become the oddballs 
which have the quantum number $J^{PC}$ that $q\bq$ 
can not have, and thus can not mix with the quarkoniums \cite{coyne,prd15}. Obviously these low-lying oddballs become very important for us to search for the pure chromoballs.

Independent of the details, however, we emphasize 
the clarity of the mixing mechanism in our quark and 
chromon model. All terms in the mass matrix have clear 
physical meaning. For example we can draw the Feynman 
diagram which represents the parameter $\nu$ in 
(\ref{mm1}), and could in principle calculate it theoretically. 

Before we close this section it is worth comparing our model 
with the so-called ``model independent" calculations in 
the conventional QCD. First, let us compare our model with 
the QCD sum rule approach which use the gauge invariant 
current operator to calculate the glueball mass, which has been asserted to be model independent \cite{shif}. Here they calculate the mass of the scalar glueball from 
the simplest gauge invariant $0^{++}$ current operator 
$\langle \F_\mn \cdot \F_\mn \rangle$ which supposedly describes the glueball made of two gluons. Similarly, 
for the $2^{++}$ glueball they have
$\langle \F_{\mu \alpha} \cdot \F_{\alpha \nu} \rangle$. 
But notice that actually these operators contain two, three, 
and four gluons, so that it is difficult to justify them as two 
gluon states. 

On the other hand in our model the simplest $0^{++}$ 
and $2^{++}$ current operators are just two chromon 
states given by $\langle \X_\mu \cdot \X_\mu \rangle$ 
and $\langle \X_\mu \cdot \X_\nu \rangle$ \cite{prd80,prl81,prd81}. Similarly, for the glueballs 
made of three chromons we have 
$\langle d_{abc} X_\mu^a X_\nu^b X_\rho^c \rangle$ 
and $\langle f_{abc} X_\mu^a X_\nu^b X_\rho^c \rangle$. 
This is simply impossible in the conventional QCD. 

Exactly the same thing can be said to the lattice calculation. 
Here again the calculations are often claimed to be ``model 
independent". However, once we understand the hidden 
structures of QCD we have much simpler ways to calculate 
the physical quantities. So the conventional ``model 
independent" calculations simply become obsolete and old fashioned after the Abelian decomposition provides new and simpler ways to calculate the physical quantities. This is 
the advantage of the Abelian decomposition. 

\section{Numerical Analysis}

The above discussion shows that the mixing analysis 
is a crucial step for us to identify the glueballs. For 
the $3\times 3$ mixing the mass matrix has five parameters, 
but we can fix $E$ and $\Delta$ from the $q\bq$ flavour 
octet data. So we need three inputs to fix the mass matrix 
completely. There are different ways to fix them. One 
way is to choose two mass eigenstates from PDG and treat 
$G$ (or equivalently the chromon mass $\mu$) as a free 
parameter, and find the best fit for $\mu$ which could 
predict the third physical state and explain the PDG data
best. Another way is to use all three mass eigenstates
as the input, and determine the the chromon mass as well. 

For the $4\times 4$ mixing the matrix has seven parameters, but we can reduce this number to five 
fixing two of them from the $q\bq$ flavor octet data. 
With this we may choose four mass eigenstates as 
the input (when available) and find the physical 
contents of the mass eigenstates, treating the chromon mass as free parameters. Or we may choose three mass eigenstates as the input and predict the mass of 
the fourth physical state, imposing an extra constraint, 
e.g., $\nu'=\nu$ etc.

In the preceding paper we have discussed the numerical 
analysis of the mixing below 2 GeV in $0^{++},~2^{++}$ 
and $0^{-+}$ channels with this strategy \cite{prd15}. 
But the numerical analysis was preliminary and inconclusive, 
partly because it depends very much on how to choose 
the inputs. In the following we discuss the mixing in 
more detail, and improve the results of the preceding 
paper.  

\subsection{$0^{++}$ channel}
 
In this channel PDG lists five iso-singlet mesons, 
$f_0(500)$, $f_0(980)$, $f_0(1370)$, $f_0(1500)$, 
and $f_0(1710)$ below 2 GeV \cite{pdg}. But 
the interpretation of the scalar mesons has been 
difficult and controversial, because some of them 
have unusually large decay width and some of 
them could be viewed as non-$q\bq$ multi-quark 
states \cite{prep1,prep2,prep3}. In this paper we try to
figure out their physical content within the quark and 
chromon model with the following two important issues 
in mind. 

The first issue is what should we choose to be the isotriplet 
partner of the flavor octet in this channel. This is very 
important because this determines the inputs $E$ and 
$\Delta$ in the mixing analysis. PDG suggests that 
the flavour octet partner of the $0^{++}$ isosinglet are 
$a_0(1450)$ and $K_0^*(1430)$, not $a_0(980)$ and 
$K_0^*(1430)$ \cite{pdg}. Intuitively, this looks somewhat 
strange because this implies that the $u$ and $d$ quarks
are heavier (or at least not lighter) than the $s$ quark. 
So it is worth for us to study the possibility that $a_0(980)$ 
and $K_0^*(1430)$ become the octet partners. 

If we adopt the PDG view and identify $a_0(1450)$ to be 
the isotriplet partner, we may choose \cite{pdg}
\bea
&E=m^2(a_0),~~~~a_0=a_0(1450), \nn \\
&\Delta =2(m^2(K)-m^2(a_0)),~~~~K=K_0^*(1430).
\label{0++1}
\eea
as the input. But as we have remarked, it is worth for us 
to check whether this PDG view is correct or not. Since 
the strange meson of the flavour octet of this channel 
is $K_0^*(1430)$, one would expect the mass of 
the non-strange isotriplet partner to be less than $1430$ 
MeV. 

In this case $a_0(980)$ becomes a natural candidate 
of the isotriplet partner of the flavour octet, and we 
may choose 
\bea
&E=m^2(a_0), ~~~~a_0=a_0(980), \nn \\
&\Delta =2(m^2(K)-m^2(a_0)),~~~~K=K_0^*(1430).
\label{0++2}
\eea
as the input. So we have two possible inputs, (\ref{0++1}) 
and (\ref{0++2}).

The second issue is the interpretation of $f_0(500)$, 
which has an unusually broad decay width. According 
to PDG it does not fit to the quark model well, and 
there have been suggestions that it could be either 
a tetra-quark state or a mixed state \cite{f5001,f5002,f5003,f5004,f5005,f5006,f5007}. 
But there are other logical possibilities. 

First, it could be viewed as a neuroball, the glueball made 
of neurons \cite{prd15}. As we pointed out, the neurons 
(just like the photons in QED) have very weak binding
because they can interact only through the quark or 
chromon loops. Nevertheless they could form a loosely 
bound state which has a broad decay width. And this is 
exactly what we find in $f_0(500)$. This is in line with 
the popular interpretation that $f_0(500)$ is a tetra-quark 
state \cite{f5001,f5002,f5003,f5004,f5005,f5006,f5007}. 
This is evident in Fig. (\ref{nball}), where the loops 
can be viewed as $q\bq$ or $gg$ bound states. 

Another possibility is that $f_0(500)$ could be the monoball, 
the vacuum fluctuation mode of the monopole condensation, 
in QCD \cite{prd15}. As we have pointed out, if the color 
confinement comes from the monopole condensation, 
QCD could have a $0^{++}$ vacuum fluctuation 
mode \cite{prd80,prl81}. In this case $f_0(500)$ becomes 
a natural candidate of this vacuum fluctuation. This 
suggests that $f_0(500)$ may not be a simple chromoball
or $q\bq$ state.

\begin{table*}[htbp]
\caption{\label{t0+1} The numerical analysis of 
the $3\times 3$ mixing in the $0^{++}$ channel, with $a_0(1450)$, $f_0(1500)$, and $f_0(1710)$ as the input. Here the third physical state can be identified as 
$f_0(1370)$.}
\begin{center}
\begin{tabular}{cccccccccccccccc}
\hline
\hline
$\mu$ & $A$ & $\nu$ &$m_3$ & \multicolumn{3}{c} {$m_1=f_0(1500)$} 
& \multicolumn{3}{c} {$m_2=f_0(1710)$} & \multicolumn{3}{c} {$m_3$} 
& $R(m_2/m_1)$ & $R(m_3/m_1)$\\
$~$&~&~&~& $u+d$ & $s$ & $G$ & $u+d$ & $s$ & $G$ & $u+d$ 
& $s$ & $G$ &~&~\\
\hline
0.76 & 0.27  & 0.18  & 1.40 & 0.07 & 0.00 & 0.93 & 0.73 & 0.20 
& 0.07 & 0.19 & 0.80 & 0.00& 0.05 & 0.00 \\
\hline
$0.78$& 0.23  & 0.31 & 1.40 & 0.26 & 0.01 & 0.73 & 0.59 & 0.16 
& 0.25 & 0.15 & 0.83 & 0.02& 0.14 & 0.02 \\
\hline
$0.80$& 0.18 & 0.36 & 1.39 & 0.44 & 0.01 & 0.54 & 0.45 & 0.12 
& 0.43 & 0.11 & 0.87 & 0.02& 0.59 & 0.05 \\
\hline
$0.82$& 0.14 & 0.35 & 1.39 & 0.62 & 0.02 & 0.36 & 0.30 & 0.08 
& 0.62 & 0.09 & 0.90 & 0.01& 1.26 & 0.07 \\
\hline
$0.84$& 0.09 & 0.29 & 1.39 & 0.79 & 0.02 & 0.18 & 0.15 & 0.04 
& 0.80 & 0.05 & 0.93 & 0.01& 3.26 & 0.09 \\
\hline
$0.86$& 0.04 & 0.07 & 1.39 & 0.96 & 0.03 & 0.01 & 0.01 & 0.00 
& 0.99 & 0.03 & 0.97 & 0.00& 85.71 & 0.12 \\
\hline
\hline
\end{tabular}
\end{center}
\end{table*}

\begin{table*}[htbp]
\caption{\label{t0+2} The numerical analysis of the $3\times 3$ mixing in the $0^{++}$ channel, with $a_0=a_0(980)$, $f_0(980)$, and $f_0(1500)$ as 
the input. Here the third physical state could be 
identified as $f_0(1710)$.}
\begin{center}
\begin{tabular}{cccccccccccccccc}
\hline
\hline
$\mu$&$A$&$\nu$&$m_3$ & \multicolumn{3}{c} {$m_1=f_0(980)$}
& \multicolumn{3}{c} {$m_2=f_0(1500)$} & \multicolumn{3}{c} {$m_3$}
& $R(m_2/m_1)$ & $R(m_3/m_1)$\\
$~$&~&~&~& $u+d$ & $s$ & $G$ & $u+d$ & $s$ & $G$ & $u+d$ 
& $s$ & $G$ &~&~\\
\hline
$0.55$& 2.44 & 1.29  & 3.06 & 0.05 & 0.00 & 0.95 & 0.44 & 0.54 
& 0.02 
& 0.51 & 0.46 & 0.02 & 0.01 & 0.00\\
\hline
$0.60$& 1.91 & 1.62  & 2.83 & 0.11 & 0.00 & 0.89 & 0.43 & 0.52 
& 0.05 
& 0.46 & 0.48 & 0.06 & 0.04 & 0.00\\
\hline
$0.65$& 1.33 & 1.68  & 2.55 & 0.21 & 0.00 & 0.79 & 0.40 & 0.49 
& 0.11 
& 0.39 & 0.51 & 0.10 & 0.09 & 0.01\\
\hline
$0.70$& 0.71 & 1.44  & 2.21 & 0.41 & 0.00 & 0.59 & 0.34 & 0.42 
& 0.24 
& 0.24 & 0.58 & 0.17 & 0.26 & 0.05\\
\hline
$0.75$& 0.04 & 0.36  & 1.79 & 0.95 & 0.00 & 0.05 & 0.05 & 0.06 
& 0.89 
& 0.00 & 0.94 & 0.05 & 10.75 & 0.43\\
\hline
\hline
\end{tabular}
\end{center}
\end{table*}

With this in mind, we can discuss the mixing. Let us first 
exclude $f(500)$ in the mixing for the reason discussed above. In the preceding paper we have discussed the $3\times 3$ mixing with $f_0(1500)$ and $f_0(1710)$ 
as the input, adopting the PDG view (\ref{0++1}) \cite{prd15}. The result is copied in Table \ref{t0+1}. 
Notice that the table shown in the preceding paper 
had a typological mistake that the numbers in the last 
two columns (i.e.,$R(m_2/m_1)$ and $R(m_3/m_1)$) 
were interchanged. This mistake is corrected in 
Table \ref{t0+1}.

The table shows that the third state (with mass around 
1400 MeV) could be identified to be $f_0(1370)$, which becomes predominantly an $s\bar s$ state.  But 
the physical contents of two other states depend very 
much on the value of the chromon mass parameter $\mu$. When $\mu$ is around 760 MeV, $f_0(1500)$ become predominantly a chromoball state and $f_0(1710)$ 
becomes predominantly the $u\bu+d\bd$ state. 

On the other hand, when $\mu$ increases to 860 MeV, 
$f_0(1500)$ becomes a $u\bu+d\bd$ state and 
$f_0(1710)$ quickly becomes a chromoball state.  But $f_0(1370)$ remains to be the $s\bs$ state, so that 
here the $s\bs$ state becomes lighter than 
the $u\bu+d\bd$ state. This, of course, is due to 
the input (\ref{0++1}). This is against the PDG 
interpretation, which suggests that $f_0(1370)$ 
is the $u\bu+d\bd$ state and $f_0(1710)$ is the 
$s\bs$ state. 

On the other hand if we adopt (\ref{0++2}) as the input, 
we obtain Table \ref{t0+2}. Here we have chosen 
$f_0(980)$ and $f_0(1500)$ as the input.  The result 
shows that when $\mu\simeq$ 750 MeV, the third state 
has mass around 1800 MeV and could be identified as $f_0(1710)$. In this case $f_0(1500)$ remains 
predominantly a chromoball state, but $f_0(1710)$ 
becomes predominantly the $s\bs$ state and $f_0(980)$ becomes predominantly the $u\bu+d\bd$ state. This of course is what we have expected from (\ref{0++2}).   

\begin{table*}[htbp]
\caption{\label{t0+3} The numerical analysis of the $4\times4$ mixing in the $0^{++}$ channel, with 
$f_0(980)$, $f_0(1370)$, $f_0(1500)$, and $f_0(1710)$ 
as the input. Here $a_0(1450)$ is identified as 
the isotriplet partner.}
\begin{center}
\begin{tabular}{cccccccccccccccccc}
\hline
\hline
$\mu$  & \multicolumn{4}{c} {$m_1=f_0(980)$}
& \multicolumn{4}{c} {$m_2=f_0(1370)$} 
& \multicolumn{4}{c} {$m_3=f_0(1500)$}
& \multicolumn{4}{c} {$m_4=f_0(1710)$} \\
$~$&  $u+d$ & $s$ & $G$ &  $G'$ &$u+d$ & $s$ & $G$& $G'$ 
& $u+d$ & $s$ & $G$&  $G'$ & $u+d$ & $s$ & $G$ & $G'$ \\
\hline
$0.50$  & 0.01 & 0.01 & 0.99 & 0.00 & 0.19 & 0.81 & 0.00 & 0.01 
& 0.13 & 0.00 & 0.00 & 0.86 & 0.68 & 0.18 & 0.01 & 0.13 \\
\hline
$0.52$  & 0.03 & 0.03 & 0.94 & 0.00 & 0.18 & 0.80 & 0.01 & 0.01 
& 0.17 & 0.01 & 0.02 & 0.81 & 0.62 & 0.17 & 0.04 & 0.17 \\
\hline
$0.54$  & 0.06 & 0.05 & 0.88 & 0.01 & 0.18 & 0.79 & 0.01 & 0.01 
& 0.21 & 0.01 & 0.04 & 0.74 & 0.55 & 0.15 & 0.06 & 0.24 \\
\hline
$0.56$  & 0.09 & 0.08 & 0.82 & 0.01 & 0.18 & 0.79 & 0.02 & 0.01 
& 0.26 & 0.01 & 0.08 & 0.65 & 0.47 & 0.13 & 0.08 & 0.33 \\
\hline
$0.58$  & 0.13 & 0.11 & 0.75 & 0.01 & 0.18 & 0.78 & 0.03 & 0.01 
& 0.32 & 0.01 & 0.14 & 0.53 & 0.38 & 0.10 & 0.08 & 0.44 \\
\hline
$0.60$  & 0.17 & 0.14 & 0.67 & 0.02 & 0.17 & 0.77 & 0.05 & 0.01 
& 0.38 & 0.01 & 0.22 & 0.39 & 0.27 & 0.07 & 0.07 & 0.58 \\
\hline
\hline
\end{tabular}

\begin{tabular}{cccccccc}
\hline
\hline
 $\mu$  & $R(m_2/m_1)$ & $R(m_3/m_1)$
 & $R(m_4/m_1)$ & A & $\nu$  & $\nu'$ & $\delta$ \\
\hline
$0.50$  & 0.01  & 0.54 & 0.06 & 0.25 & 0.19  & 0.24 & 1.35  \\
\hline
$0.52$  & 0.01  & 0.54 & 0.10 & 0.21 & 0.41  & 0.28 & 1.30  \\
\hline
$0.54$  & 0.02  & 0.54 & 0.16 & 0.17 & 0.53  & 0.32 & 1.26  \\
\hline
$0.56$  & 0.03  & 0.54 & 0.22 & 0.13 & 0.61  & 0.36 & 1.22  \\
\hline
$0.58$  & 0.04  & 0.54 & 0.31 & 0.07 & 0.66  & 0.40 & 1.20  \\
\hline
$0.60$  & 0.06  & 0.54 & 0.43 & 0.01 & 0.68  & 0.42 & 1.21  \\
\hline
\hline
\end{tabular}
\end{center}
\end{table*}

\begin{table*}[htbp]
\caption{\label{t0+4} The numerical analysis of the $4\times4$ mixing in the $0^{++}$ channel, with 
$f_0(980)$, $f_0(1370)$, $f_0(1500)$, and $f_0(1710)$ 
as the input. Here $a_0(980)$ is identified as 
the isotriplet partner.}
\begin{center}
\begin{tabular}{cccccccccccccccccc}
\hline
\hline
$\mu$  & \multicolumn{4}{c} {$m_1=f_0(980)$}
& \multicolumn{4}{c} {$m_2=f_0(1370)$} 
& \multicolumn{4}{c} {$m_3=f_0(1500)$}
& \multicolumn{4}{c} {$m_4=f_0(1710)$} \\
$~$&  $u+d$ & $s$ & $G$ &  $G'$ &$u+d$ & $s$ & $G$& $G'$ 
& $u+d$ & $s$ & $G$& $G'$ & $u+d$ & $s$ & $G$& $G'$ \\
\hline
$0.67$  & 0.86 & 0.00 & 0.14 & 0.01 & 0.13 & 0.04 & 0.83 & 0.04 
& 0.01 & 0.01 & 0.00 & 0.98 & 0.01 & 0.96 & 0.03 & 0.01 \\
\hline
$0.68$  & 0.90 & 0.00 & 0.05 & 0.05 & 0.04 & 0.01 & 0.94 & 0.01 
& 0.05 & 0.06 & 0.00 & 0.89 & 0.01 & 0.93 & 0.01 & 0.05 \\
\hline
\hline
\end{tabular}

\begin{tabular}{cccccccc}
\hline
\hline
 $\mu$  & $R(m_2/m_1)$ & $R(m_3/m_1)$
 & $R(m_4/m_1)$ & A & $\nu$  & $\nu'$ & $\delta$\\
\hline
$0.67$   & 4.17  & 4.21 & 0.11 & 0.08 & 0.40  & 0.13 & 0.47  \\
\hline
$0.68$   & 7.07  & 5.73 & 0.26 & 0.07 & 0.25  & 0.35 & 0.40  \\
\hline
\hline
\end{tabular}
\end{center}
\end{table*}

Clearly the two tables give different descriptions, 
and we have to know which is closer to the truth. 
One way to find which is better is to compare 
the predictions of the relative radiative decay ratios 
of $J/\psi$ with the experimental data. Experimentally 
PDG has new data on the radiative decay of $J/\psi$ to 
$0^{++}$ states \cite{pdg},
\bea
&J/\psi\rightarrow\gamma f_0(1710) \rightarrow
\left\{\begin{array}{ll} \gamma K\bar K 
\simeq (8.5+1.2-0.9) \\
\times 10^{-4},  \\
\gamma \pi\pi \simeq (4.0\pm 1.0) \times 10^{-4},  \\
\gamma \omega \omega \simeq (3.1\pm 1.0) \times 10^{-4},  \\
\gamma \eta \eta  \simeq (2.4+ 1.2-0.7) \\
\times 10^{-4}, 
\end{array}\right. \nn
\eea
and
\bea
&J/\psi\rightarrow\gamma f_0(1500)\rightarrow
\left\{\begin{array}{ll} \gamma \pi\pi
\simeq (1.01\pm 0.32) \times 10^{-4},   \\
\gamma \eta\eta \simeq (1.7+ 0.6-1.4) \\
\times 10^{-5}. 
\end{array}\right.   \nn
\eea
Of course this may not be the final data, because other 
decay modes could be discovered later. But assuming that 
this is the final result, we have
\bea
R(f_0(1710)/f_0(1500))\simeq 15.3.
\label{pdg0}
\eea
This is a very important piece of information, because this 
determines the glue content of the mass eigenstates. 

Now, Table \ref{t0+2} predicts the relative radiative 
decay ratio to be $R(f_0(1710)/f_0(1500)) \simeq$ 
0.04 (at $\mu=$ 750 MeV). Clearly this is not in line 
with (\ref{pdg0}), which is troublesome. In comparison, 
according to Table \ref{t0+1} we have 
$R(f_0(1710)/f_0(1500)) \simeq$ 15.3 (at $\mu \simeq$ 
856 MeV). This is in good agreement with the PDG data, which implies that Table \ref{t0+1} is better. This in turn implies that $a_0(1450)$, not $a_0(980)$, could be 
the isotriplet partner of these isosinglet states. But this conclusion is premature because the contents of physical states in Table \ref{t0+1} is controversial and 
$R(m_2/m_1)$ becomes very sensitive to the change 
of $\mu$. 

Independent of whether this result is correct or not, 
however, the above $3\times 3$ mixing has a critical 
shortcoming in that it can explain the mixing of only 
three physical states, while here we have at least four 
physical states (excluding $f_0(500)$) below 2 GeV. 
This strongly motivates us to go to the $4\times 4$ 
mixing. And this is independent of which input, 
(\ref{0++1}) or (\ref{0++2}), we choose.

So we consider the $4\times 4$ mixing (\ref{mm2})
with two chromoball states $^1S_0$ and $^5D_0$ 
made of two chromons $|G\rangle$ and $|G'\rangle$.
Diagonalizing the two chromoball mass matrix first, we 
may put 
\bea
\epsilon=0,~~~G=4\mu^2,~~~G'=G+\delta,
\eea 
and consider the mixing of the two $q\bq$ states with 
two chromoballs which have mass $\sqrt G$ and 
$\sqrt {G'}$. This has seven parameters, but we can 
fix two with (\ref{0++1}) or (\ref{0++2}) and four with 
the four mass eigenstates $f_0(980)$, $f_0(1370)$, $f_0(1500)$, and $f_0(1710)$ as the input. With this 
we can diagonalize the mass matrix and find the physical contents of the mass eigenstates, treating the chromon mass $\mu$ as the free parameter. 

Now, adopting the PDG view (\ref{0++1}) we obtain 
Table \ref{t0+3}, but with (\ref{0++2}) we obtain Table
\ref{t0+4}. But the mathematical equations which we need 
to solve to diagonalize the mass matrix are very rigid 
which often have no solution, and this forces us to change 
the input data slightly to find the solutions. So here
we have changed the four mass eigenstates to 990, 1400, 
1505, and 1722 MeVs to obtain Table \ref{t0+3}, and to 990, 
1370, 1505, 1800 MeVs to obtain Table \ref{t0+4}.
  
The numerical result of Table \ref {t0+3} obtained with
(\ref{0++1}) suggests that $f_0(980)$ is predominantly 
the $^1S_0$ chromoball state and $f_0(1370)$ is 
predominantly the $s\bs$ state. But $f_0(1500)$ becomes 
largely the $^5D_0$ chromoball state and $f_0(1710)$ 
becomes largely the $u\bu+d\bd$ state, although they 
have considerable mixing as the chromon mass increases 
to 600 MeV. This is in line with Table \ref{t0+1}. But 
here the $u\bu+d\bd$ state remains heavier than 
the $s\bs$ state, which again is due to the input 
(\ref{0++1}).

On the other hand, Table \ref{t0+4} obtained with 
(\ref{0++2}) tells that $f_0(980)$ and $f_0(1710)$ 
are the $u\bu+d\bd$ and $s\bs$ states, respectively. 
And $f_0(1370)$ and $f_0(1500)$ become the $^1S_0$ 
and $^5D_0$ chromoball states. Again this is 
consistent with Table \ref{t0+2}.

As for the $J/\psi$ radiative decay branching ratio, 
Table \ref{t0+3} shows that $R(f_0(1710)/f_0(1500))
\simeq$ 0.8 when $\mu=0.60$, and Table \ref{t0+4}
gives around 0.05 when $\mu=$ 680 MeV. Clearly both 
are too small compared to (\ref{pdg0}), so that we 
can not tell which is the isotriplet partner of 
the $0^{++}$ isosinglet state. 

The contrast between Table \ref{t0+3} and Table 
\ref{t0+4} is unmistakable. This, of course, originates 
from the inputs (\ref{0++1}) and (\ref{0++2}). This 
analysis has both positive and negative sides. 
The positive side is that the result of the $4\times 4$ 
mixing, in particular the physical contents of the mass 
eigenstates, is consistent with the $3\times 3$ mixing
analysis. But the disappointing point is that 
the $4\times 4$ analysis can not tell whether the PDG 
view that $a_0(1450)$, not $a_0(980)$, is the isotriplet 
partner of the $0^{++}$ isosinglet state. 

\begin{table*}[htbp]
\caption{\label{t0+5} The numerical analysis of the $5\times5$ mixing in the $0^{++}$ channel, with all
five mass eigenstates ($f_0(500)$, $f_0(980)$, 
$f_0(1370)$, $f_0(1500)$, and $f_0(1710)$) as the input. Here $a_0(1450)$ is identified as the isotriplet partner.}
\begin{center}
\begin{tabular}{cccccccccccccccccccccccccccccc}
\hline
\hline
$\mu_0$  & $\mu$  & \multicolumn{5}{c} {$m_1=f_0(500)$}
& \multicolumn{5}{c} {$m_2=f_0(980)$} & \multicolumn{5}{c} {$m_3=f_0(1370)$} 
& \multicolumn{5}{c} {$m_4=f_0(1500)$} \\
$~$&~& $u+d$ & $s$ & $G_0$ & $G_1$ & $G_2$ &$u+d$ & $s$ & $G_0$ 
& $G_1$ & $G_2$ & $u+d$ & $s$ & $G_0$ & $G_1$& $G_2$ & $u+d$ 
& $s$ & $G_0$ & $G_1$ & $G_2$ \\
\hline
$0.28$ & 0.50 & 0.00 & 0.00 & 1.00 & 0.00 & 0.00 & 0.00 & 0.00 & 0.00 & 1.00 
& 0.00 & 0.18 & 0.81 & 0.00 & 0.00 & 0.01 & 0.13 & 0.00 & 0.00 & 0.00 & 0.87 \\
\hline
$0.30$ & 0.52 & 0.01 & 0.01 & 0.97 & 0.00 & 0.00 & 0.03 & 0.02 & 0.01 & 0.94 
& 0.00 & 0.18 & 0.80 & 0.00 & 0.01 & 0.01 & 0.18 & 0.01 & 0.01 & 0.02 & 0.79 \\
\hline
$0.32$ & 0.53 & 0.03 & 0.02 & 0.92 & 0.02 & 0.00 & 0.04 & 0.04 & 0.05 & 0.87 
& 0.00 & 0.18 & 0.79 & 0.00 & 0.01 & 0.01 & 0.23 & 0.01 & 0.01 & 0.04 & 0.71 \\
\hline
$0.34$ & 0.55 & 0.05 & 0.04 & 0.86 & 0.04 & 0.00 & 0.06 & 0.05 & 0.09 & 0.80 
& 0.01 & 0.18 & 0.79 & 0.00 & 0.02 & 0.01 & 0.27 & 0.01 & 0.02 & 0.08 & 0.63 \\
\hline
$0.36$ & 0.56 & 0.08 & 0.05 & 0.80 & 0.07 & 0.00 & 0.06 & 0.05 & 0.15 & 0.73 
& 0.01 & 0.18 & 0.78 & 0.00 & 0.02 & 0.01 & 0.31 & 0.01 & 0.02 & 0.11 & 0.55 \\
\hline
\hline
\end{tabular}
\begin{tabular}{ccccccccccccccccccc}
\hline
\hline
$\mu_0$ & $\mu$ & \multicolumn{5}{c} {$m_5=f_0(1710)$} & $R(m_2/m_1)$ 
& $R(m_3/m_1)$ & $R(m_4/m_1)$ & $R(m_5/m_1)$ & A & $\nu_1$ & $\nu_2$  
& $\delta$ & $\delta'$ \\
~&~& $u+d$ & $s$ & $G_0$ & $G_1$ & $G_2$ &~&~&~&~&~&~&~&~&~&~\\
\hline
$0.28$ & 0.50 & 0.69 & 0.19 & 0.00 & 0.00 & 0.12 & 0.80 & 0.01 & 0.43 
& 0.05 & 0.26 & 0.10 & 0.23 & 0.67 & 2.03 \\
\hline
$0.30$ & 0.52 & 0.60 & 0.16 & 0.01 & 0.03 & 0.19 & 0.78 & 0.01 & 0.41 
& 0.09 & 0.20 & 0.29 & 0.40 & 0.71 & 2.09 \\
\hline
$0.32$ & 0.53 & 0.52 & 0.14 & 0.02 & 0.06 & 0.27 & 0.77 & 0.02 & 0.40 
& 0.13 & 0.14 & 0.54 & 0.34 & 0.73 & 2.03 \\
\hline
$0.34$ & 0.55 & 0.44 & 0.12 & 0.02 & 0.08 & 0.35 & 0.79 & 0.02 & 0.39 
& 0.17 & 0.09 & 0.61 & 0.38 & 0.74 & 2.03 \\
\hline
$0.36$ & 0.56 & 0.38 & 0.10 & 0.02 & 0.07 & 0.42 & 0.81 & 0.03 & 0.39 
& 0.22 & 0.04 & 0.67 & 0.40 & 0.73 & 2.02 \\
\hline
\hline
\end{tabular}
\end{center}
\end{table*}

\begin{table*}[htbp]
\caption{\label{t0+6} The numerical analysis of the $5\times5$ 
mixing in the $0^{++}$ channel, with all 5 mass eigenstates 
($f_0(500)$, $f_0(980)$, $f_0(1370)$, $f_0(1500)$, and 
$f_0(1710)$) as the input. Here $a_0(980)$ is identified 
as the isotriplet partner.}
\begin{center}
\begin{tabular}{cccccccccccccccccccccccccccccc}
\hline
\hline
$\mu_0$ &$\mu$ &  \multicolumn{5}{c} {$m_1=f_0(500)$}
& \multicolumn{5}{c} {$m_2=f_0(980)$} & \multicolumn{5}{c} {$m_3=f_0(1370)$} 
& \multicolumn{5}{c} {$m_4=f_0(1500)$} \\
$~$&~& $u+d$ & $s$ & $G_0$ & $G_1$& $G_2$ & $u+d$ & $s$ & $G_0$ 
& $G_1$ & $G_2$ & $u+d$ & $s$ & $G_0$ & $G_1$ & $G_2$ & $u+d$ 
& $s$ & $G_0$ & $G_1$ & $G_2$ \\
\hline
$0.28$ & 0.66 & 0.01 & 0.00 & 0.99 & 0.00 & 0.00 & 0.85 & 0.00 & 0.01 & 0.12 
& 0.01 & 0.12 & 0.03 & 0.00 & 0.83 & 0.01 &0.01 & 0.02 & 0.00 & 0.00 & 0.97 \\
\hline
$0.30$ & 0.69 & 0.07 & 0.00 & 0.92 & 0.00 & 0.00 & 0.85 & 0.00 & 0.07 & 0.00 
& 0.08 & 0.00 & 0.00 & 0.00 & 1.00& 0.00 & 0.08 & 0.09 & 0.00 & 0.00 & 0.83 \\
\hline
$0.32$ & 0.68 & 0.13 & 0.00 & 0.86 & 0.00 & 0.01 & 0.77 & 0.00 & 0.14 & 0.01 
& 0.08 & 0.00 & 0.00 & 0.00 & 1.00 & 0.00 & 0.08 & 0.10 & 0.00 & 0.00 & 0.82 \\
\hline
$0.34$ & 0.68 & 0.20 & 0.01 & 0.78 & 0.00 & 0.01 & 0.69 & 0.00 & 0.21 & 0.02 
& 0.08 & 0.02 & 0.00 & 0.00 & 0.97 & 0.01 & 0.09 & 0.11 & 0.00 & 0.00 & 0.81 \\
\hline
$0.36$ & 0.68 & 0.26 & 0.01 & 0.71 & 0.01 & 0.02 & 0.61 & 0.00 & 0.28 & 0.03 
& 0.07 & 0.03 & 0.01 & 0.00 & 0.95 & 0.02 & 0.09 & 0.11 & 0.00 & 0.00 & 0.80 \\
\hline
\hline
\end{tabular}
\begin{tabular}{ccccccccccccccccccc}
\hline
\hline
$\mu_0$ & $\mu$ & \multicolumn{5}{c} {$m_5=f_0(1710)$} &$R(m_2/m_1)$
& $R(m_3/m_1)$ & $R(m_4/m_1)$& $R(m_5/m_1)$ & A & $\nu_1$ & $\nu_2$  
& $\delta$ & $\delta'$ \\
~&~& $u+d$ & $s$ & $G_0$ & $G_1$ & $G_2$ &~&~&~&~&~&~&~&~&~&~\\
\hline
$0.28$ & 0.66 & 0.01 & 0.95 & 0.00 & 0.03 & 0.01 & 0.12 & 0.49 & 0.48 
& 0.01 & 0.08 & 0.39 & -0.18 & 1.48 & 1.95 \\
\hline
$0.30$ & 0.69 & 0.01 & 0.90 & 0.00 & 0.01 & 0.09 & 0.13 & 0.62 & 0.44 
& 0.03 & 0.04 & 0.02 & -0.48 & 1.52 & 1.88 \\
\hline
$0.32$ & 0.68 & 0.01 & 0.89 & 0.00 & 0.00 & 0.09 & 0.21 & 0.66 & 0.46 
& 0.04 & 0.03 & 0.50 & 0.10 & 1.46 & 1.83 \\
\hline
$0.34$ & 0.68 & 0.01 & 0.88 & 0.01 & 0.01 & 0.10 & 0.31 & 0.70 & 0.50 
& 0.04 & 0.01 & 0.19 & 0.51 & 1.40 & 1.77 \\
\hline
$0.36$ & 0.68 & 0.01 & 0.87 & 0.01 & 0.01 & 0.10 & 0.42 & 0.76 & 0.54 
& 0.05 & 0.00 & 0.25 & 0.53 & 1.34 & 1.71 \\
\hline
\hline
\end{tabular}
\end{center}
\end{table*}

So far we have excluded $f(500)$ in the mixing, because 
it could be viewed as a neuroball or monoball, not a chromoball. On the other hand, there is no reason why 
it can not mix with quarkoniums and chromoballs. 
Actually, even when $f_0(500)$ becomes a neuroball 
it makes sense to include it in the mixing, because 
the neuroball could be viewed as a glueball. This must 
be clear from Fig. \ref{nball} and Fig. \ref{mixing}. 
Moreover, even when it becomes the monoball, 
the vacuum fluctuation of the monopole condensation, 
there is no reason why it could not mix with quarkoniums and chromoballs. This justifies the $5\times 5$ mixing. 

For this reason we consider the following $5\times 5$ 
mixing with two $q\bq$ and three glueballs which has 
nine parameters,
\bea
M^2 =\left( \begin{array}{ccccc}
E+\dfrac{2}{3}\Delta  & -\dfrac{\sqrt{2}}{3} \Delta & 
0 & 0 &0 \\
-\dfrac{\sqrt{2}}{3}\Delta  & E+\dfrac{1}{3}\Delta 
+3A & \nu_0 & \nu_1 & \nu_2 \\
0  &  \nu_0 & G_0 & 0 & 0  \\
0  & \nu_1 & 0 & G_1 & 0  \\
0  & \nu_2 & 0 & 0 & G_2 \end{array} \right),
\label{massm4}
\eea 
Here $G_1$ and $G_2$ are the $^1S_0$ and $^5D_0$ 
chromoball as before, but $G_0$ is supposed to be
the monoball or the neuroball. 

In this case we can put all five physical states below 
2 GeV, including $f_0(500)$, and adopt (\ref{0++1}) 
or (\ref{0++2}) as the input, and treat $\mu$ as a free 
parameter. But we need to impose one more constraint 
to fix the mass matrix completely.
  
To do that we may have to take into account the possibility 
that $G_0$ is not an ordinary chromoball. There are two 
possibilities. If $f_0(500)$ is the neuroball, it is natural to 
expect $\nu_0$ to be of the same order as $\nu_1$ and 
$\nu_2$. This must be clear from the mixing diagram 
Fig. \ref{mixing}. In this case we could assume 
\bea
\nu_0\simeq \dfrac{\nu_1+\nu_2}{2}.
\label{c0+55}
\eea
On the other hand, if $f_0(500)$ is the monoball, its 
coupling to ordinary chromoball and quarkonium states 
could be of second order. In this case $\nu_0$ could 
be much smaller than $\nu_1$ and $\nu_2$. Here we 
will use (\ref{c0+55}) as the constraint. Of course, we 
emphasize that there is no justification for this. We 
assume this just for simplicity to fix the mass matrix 
completely.

Now, with 
\bea
&G_0=4\mu_0^2,~~~~~G_1=G_0+\delta=4\mu^2,   \nn\\
&G_2=G_0+\delta',
\eea
we obtain Table \ref{t0+5} using (\ref{0++1}), and 
Table \ref{t0+6} using (\ref{0++2}). But diagonalizing 
the $5\times 5$ matrix involves solving a sixth order 
polynomial, and it is not easy to find the solution with 
the input data. So we have changed the input masses 
a little bit, and used 550, 990, 1400, 1505, and 
1722 MeVs for $f_0(500)$, $f_0(980)$, $f_0(1370)$, 
$f_0(1500)$, and $f_0(1710)$ to obtain Table \ref{t0+5}, 
and 550, 990, 1370, 1505, and 1800 for $f_0(500)$, 
$f_0(980)$, $f_0(1370)$, $f_0(1500)$, and $f_0(1710)$ 
to obtain Table \ref{t0+6}.

Notice that here we have expressed $G_0$ in terms of 
the neuron effective mass $\mu_0$ (assuming that $G_0$ 
is the neuroball) to compare it with the chromon mass 
$\mu$ fixed by $G_1$. In this $5\times 5$ mixing 
the glueballs play the dominant role, because only 
two of the physical states can be the $q\bq$ states. 
So here the issue becomes which of the five physical 
states are the $q\bq$ states, not the glueball states. 

Table \ref{t0+5} tells that $f_0(1370)$ is predominantly 
the $s\bs$ state, and $f_0(1710)$ becomes the mixed 
state about half of which is the $u\bu+d\bd$ state (and 
$f_0(1500)$ becomes the mixed states about quarter 
of which is the $u\bu+d\bd$ state). This is consistent 
with Table \ref{t0+1}. Moreover, the table tells 
the followings. First, $f_0(500)$ is the mainly the lowest 
energy glueball which could be interpreted as either 
the neuroball or the monoball. Second, $f_0(980)$ and 
$f_0(1500)$ are mainly the $^1S_0$ and $^5D_0$ chromoball 
states, and a considerable part of $f_0(1710)$ is made of 
$^5D_0$ chromoball. 

In comparison Table \ref{t0+6} tells that $f_0(980)$ and 
$f_0(1710)$ predominantly the $u\bu+d\bd$ and $s\bs$ 
states, respectively. This is in line with Table \ref{t0+2}. 
Moreover, this table tells that $f_0(500)$ is mainly the lowest 
energy the neuroball (or the monoball), and $f_0(1370)$ 
and $f_0(1500)$ are mainly the $^1S_0$ and $^5D_0$ 
chromoball states. 

In the literature there have been diverse interpretations 
of the scalar mesons. One of the popular views is that 
$f_0(500)$ and $f_0(980)$ are the tetra-quark states \cite{f5001,f5002,f5003,f5004,f5005,f5006,f5007,
f9801,f9802,f9803,f9804,f9805}, $f_0(1370)$ and 
$f_0(1500)$ are the mixed state \cite{f13701,f13702,
f13703,f13704,f13705,f15001,f15002}, and $f_0(1710)$ 
is a scalar glueball \cite{f17101,f17102,f17103}. Another 
view is that $f_0(1370)$, $f_0(1710)$, $a_0(1450)$, 
and $K_0^*(1430)$ are the members of the flavor nonet, 
$f_0(1710)$ being mainly the $s\bs$ state \cite{ams,close}. 
In this view $f_0(1500)$ can naturally be identified as 
predominantly the glueball state. And this seems to be 
endorsed by PDG \cite{pdg}.

\begin{table*}[htbp]
\caption{\label{t2+1}  The numerical analysis of 
the $3\times 3$ mixing in the $2^{++}$ channel, with $f_2(1270)$, $f'_2(1525)$ and $f_2(1950)$ as the input.}
\begin{center}
\begin{tabular}{ccccccccccccccc}
\hline
\hline
$\mu$ & \multicolumn{3}{c} {$m_1=f_2'(1270)$} 
& \multicolumn{3}{c} {$m_2=f_2'(1525)$} 
& \multicolumn{3}{c} {$m_3=f_2(1950)$} & $R(m_2/m_1)$ 
& $R(m_3/m_1)$ & $A$ & $\nu$ \\
$~$ & $u+d$ & $s$ & $G$ & $u+d$ & $s$ & $G$ 
& $u+d$ & $s$ & $G$ &~&~&~&~ \\
\hline
$0.92$ & 0.86 & 0.01 & 0.13 & 0.04 & 0.88 & 0.07 
& 0.10 & 0.10 & 0.80& 0.41 & 2.33 & 0.07 & 0.79 \\
\hline
\hline
\end{tabular}
\end{center}
\end{table*}

\begin{table*}[htbp]
\caption{\label{t2+2} The numerical analysis of the $4\times4$ 
mixing in the $2^{++}$ channel, with $f_2(1270)$, $f_2'(1525)$, 
$f_2(1950)$, and $f_2(2010)$ as the input.}
\begin{center}
\begin{tabular}{cccccccccccccccccccccc}
\hline
\hline
$\mu$ & \multicolumn{4}{c} {$m_1=f_2(1270)$}
& \multicolumn{4}{c} {$m_2=f_2'(1525)$} 
& \multicolumn{4}{c} {$m_3=f_2(1950)$} 
& \multicolumn{4}{c} {$m_4=f_2(2010)$} \\
$~$ & $u+d$ & $s$ & $G$ & $G'$ &$u+d$ & $s$ & $G$ & $G'$ 
& $u+d$ & $s$ & $G$ & $G'$ & $u+d$ & $s$ & $G$ & $G'$ \\
\hline
$0.90$ & 0.80 & 0.01 & 0.19 & 0.00 & 0.06 & 0.85 & 0.09 & 0.00 
& 0.13 & 0.14 & 0.72 & 0.01 & 0.01 & 0.00 & 0.01 & 0.99 \\
\hline
$0.91$ & 0.79 & 0.01 & 0.18 & 0.02 & 0.06 & 0.85 & 0.08 & 0.01 
& 0.09 & 0.09 & 0.67 & 0.15 & 0.05 & 0.05 & 0.07 & 0.83 \\
\hline
$0.92$ & 0.79 & 0.01 & 0.17 & 0.03 & 0.06 & 0.85 & 0.07 & 0.01 
& 0.06 & 0.06 & 0.61 & 0.27 & 0.09 & 0.08 & 0.14 & 0.69 \\
\hline
$0.93$ & 0.79 & 0.01 & 0.16 & 0.04 & 0.06 & 0.86 & 0.07 & 0.01 
& 0.03 & 0.03 & 0.54 & 0.39 & 0.12 & 0.10 & 0.22 & 0.55 \\
\hline
$0.94$ & 0.78 & 0.01 & 0.16 & 0.04 & 0.06 & 0.86 & 0.07 & 0.02 
& 0.01 & 0.01 & 0.44 & 0.53 & 0.14 & 0.12 & 0.33 & 0.41\\
\hline
\hline
\end{tabular}
\begin{tabular}{cccccccc}
\hline
\hline
$\mu$ & $R(m_2/m_1)$ & $R(m_3/m_1)$
& $R(m_4/m_1)$ & A & $\nu$ & $\nu'$ & $\delta$ \\
\hline
$0.90$ & 1.49 & 1.43 & 0.37 & 0.11 & 0.89 & 0.14 & 1.16 \\
\hline
$0.91$ & 1.62 & 1.26 & 0.34 & 0.14 & 0.91 & 0.44 & 0.94 \\
\hline
$0.92$ & 1.71 & 1.13 & 0.33 & 0.15 & 0.93 & 0.56 & 0.75 \\
\hline
$0.93$ & 1.78 & 1.04 & 0.31 & 0.17 & 0.96 & 0.62 & 0.56 \\
\hline
$0.94$ & 1.82 & 0.98 & 0.31 & 0.18 & 0.99 & 0.62 & 0.39 \\
\hline
\hline
\end{tabular}
\end{center}
\end{table*}

Our analysis does not entirely support this. If we identify
$a_0(1450)$ as the isotriplet partner, $f_0(1370)$ and 
$f_0(1710)$ become the $q\bq$ states. But according to
Table \ref{t0+5}, $f_0(1370)$ turns out to be predominantly 
the $s\bs$ state. On the other hand Table \ref{t0+6} 
shows that  $f_0(1710)$ becomes predominantly 
the $s\bs$ state, if we identify $a_0(980)$ as the isotriplet 
partner. So, at this point it is premature to make a definite 
conclusion on which state, $a_0(980)$ or $a_0(1450)$, 
is the isotriplet partner of the $0^{++}$ isosinglet $q\bq$ 
state. We just remark that here our analysis does show 
that the possibility that $a_0(980)$ could be the isotriplet 
partner remains an option.

But we like to emphasize two remarkable results of
our mixing. First,  both tables seem to be consistent 
with the view that $f_0(500)$ is the neuroball. Moreover, they suggest that the neuron mass $\mu_0$ is around 
300 MeV, which is smaller than the chromon mass. 
This is interesting and reasonable. This should be 
compared with the popular view that $f_0(500)$ (and $f_0(980)$) are the tetra-quark states.  As we have 
pointed out, in our quark and chromon model 
the tetra-quark states could be interpreted as 
the glueballs made of two neurons or chromons. 
This must be clear from Fig. \ref{cball} and Fig. \ref{nball}. 
So this result is not inconsistent with the popular view 
that $f_0(500)$ is a tetra-quark state. 

Second, both tables suggest that $f_0(1500)$ could
be predominantly the chromoball state. This is also 
very interesting. On the other hand, in both tables 
the radiative decay ratio $R(f_0(1710)/f_0(1500))$ 
turns out to be too small compared to (\ref{pdg0}).
But we notice that the relative radiative decay ratios 
in general are very sensitive to the inputs, so that 
this could be due to the {\it ad hoc} constraint 
(\ref{c0+55}). 

To summarize, it is difficult to draw any conclusive 
result from the above numerical analysis. Perhaps 
one positive side of the above analysis is that 
$a_0(980)$ could still turn out to be the isotriplet 
partner of the flavour octet in this channel. Another
point is the physical content of $f_0(500)$. It has 
been a big mystery in hadron spectroscopy, 
on which a huge amount of literature exists \cite{f5008,f5009,f50010,f50011}. In this paper 
we studied the possibility that it could be interpreted 
as a neuroball. Our result appears to be consistent 
with this view. But it could also turn out to be 
the monoball, and we certainly need more analysis 
to make a definite conclusion on this. 

\subsection{$2^{++}$channel}

In this channel we have three physical states below 2 GeV, 
$f_2(1270)$, $f_2'(1525)$, and $f_2(1950)$. On the other 
hand we have to keep in mind that there is the fourth state 
$f_2(2010)$ just above 2 GeV, which could be included 
in the mixing. Another point is that PDG lists five more
unestablished states, $f_2(1430)$, $f_2(1565)$, 
$f_2(1640)$, $f_2(1810)$, and $f_2(1910)$, some of 
which could turn out to be real states. In this paper we 
will consider only the three and $f_2(2010)$ established states in the mixing analysis, but the fact that there 
are so many unestablished $2^{++}$ states implies that 
we have to be careful to analyse the mixing in this channel.

In the preceding paper we have studied the $3\times 3$ 
mixing of one chromoball and two quarkoniums, using
\bea
&E=m^2(a_2),~~~a_2=a_2(1320), \nn \\
&\Delta =2(m^2(K^*)-m^2(a_2)),  \nn\\
&K^*=K_2^*(1430),
\label{2++} 
\eea
with two physical states $f_2(1270)$ and $f_2(1950)$ 
as the input, and predicted the mass of the third 
physical state varying the chromon mass $\mu$ as 
a free parameter \cite{prd15}.  

The result suggests that, when the mass parameter 
$\mu$ is around 760 MeV, $f_2(1270)$ becomes 
a mixture of $u\bu+d\bd$ and chromoball, $f_2(1950)$ becomes a mixture of $u\bu+d\bd$, $s\bs$ and 
the chromoball, but $f'(1525)$ becomes predominantly 
the $s\bar s$ state. 

On the other hand when $\mu$ becomes around 860 MeV, $f_2(1270)$ becomes predominantly $u\bu+d\bd$ state, $f_2(1950)$ becomes predominantly the chromoball, 
and $f_2'(1525)$ remains predominantly the $s\bs$ 
state. This was in line with the PDG suggestion, which interprets $f_2(1270)$ and $f_2'(1525)$ as the $q\bq$ states \cite{pdg}. 

But now we have more experimental data on the $J/\psi$ 
radiative decay from PDG \cite{pdg}
\bea
&J/\psi\rightarrow\gamma f_2(1270)
\simeq (1.43\pm0.11)\times 10^{-3},   \nn\\
&J/\psi\rightarrow\gamma f_2'(1525)
\simeq (4.5+0.7-0.4)\times 10^{-4},   \nn\\
&J/\psi\rightarrow\gamma f_2(1950)
\simeq (7.0\pm2.2)\times 10^{-4},   \nn
\eea
which give us
\bea
&R(f_2(1525)/f_2(1270)) \simeq 0.36,   \nn\\
&R(f_2(1950)/f_2(1270)) \simeq 0.49,   \nn\\
&R(f_2(1950)/f'_2(1525)) \simeq 1.56. 
\label{pdg2}
\eea
So we could test these experimental data in our analysis.

\begin{table*}[htbp]
\caption{\label{t2+3} The numerical analysis of the $4\times4$ 
mixing in the $2^{++}$ channel, with states $f_2(1270)$, 
$f_2'(1525)$, $f_2(1950)$ as the input. The fourth state 
could be interpreted as $f_2(2010)$.}
\begin{center}
\begin{tabular}{ccccccccccccccccccccccc}
\hline
\hline
$\mu$ & $m_4$ &  \multicolumn{4}{c} {$m_1=f_2(1270)$}
& \multicolumn{4}{c} {$m_2=f_2'(1525)$} 
& \multicolumn{4}{c} {$m_3=f_2(1950)$} & \multicolumn{4}{c} {$m_4$} \\
$~$ & $~$ & $u+d$ & $s$ & $G$ & $G'$ & $u+d$ & $s$ & $G$ & $G'$ 
& $u+d$ & $s$ & $G$ & $G'$ & $u+d$ & $s$ & $G$ & $G'$ \\
\hline
$0.90$  & 2.86 & 0.80 & 0.01 & 0.19 & 0.00 & 0.06 & 0.85 & 0.09 
& 0.00 & 0.14 & 0.14 & 0.72 & 0.00 & 0.00 & 0.00 & 0.00 & 0.99 \\
\hline
$0.91$  & 2.11 & 0.79 & 0.01 & 0.18 & 0.02 & 0.06 & 0.85 & 0.08 
& 0.01 & 0.09 & 0.09 & 0.68 & 0.14 & 0.05 & 0.05 & 0.06 & 0.84 \\
\hline
$0.92$  & 2.07 & 0.79 & 0.01 & 0.18 & 0.02 & 0.06 & 0.85 & 0.08 
& 0.01 & 0.05 & 0.05 & 0.54 & 0.36 & 0.10 & 0.09 & 0.21 & 0.61 \\
\hline
$0.93$  & 2.07 & 0.79 & 0.01 & 0.18 & 0.03 & 0.06 & 0.86 & 0.07 
& 0.01 & 0.02 & 0.02 & 0.42 & 0.54 & 0.13 & 0.11 & 0.33 & 0.43 \\
\hline
$0.94$  & 2.08 & 0.78 & 0.01 & 0.17 & 0.03 & 0.06 & 0.86 & 0.07 
& 0.01 & 0.01 & 0.01 & 0.33 & 0.65 & 0.14 & 0.12 & 0.43 & 0.21\\
\hline
\hline
\end{tabular}

\begin{tabular}{ccccccccc}
\hline
\hline
$\mu$ & $m_4$& $R(m_2/m_1)$ & $R(m_3/m_1)$
& $R(m_4/m_1)$ & A & $\nu$  & $\epsilon$ \\
\hline
$0.90$  & 2.86 & 1.48 & 0.03 & 0.37 & 0.12 & 0.89 & 4.89 \\
\hline
$0.91$  & 2.11 & 1.61 & 1.23 & 0.34 & 0.14 & 0.91 & 0.98 \\
\hline
$0.92$  & 2.07 & 1.75 & 1.20 & 0.33 & 0.15 & 0.95 & 0.63 \\
\hline
$0.93$  & 2.07 & 1.82 & 1.10 & 0.32 & 0.16 & 0.98 & 0.46 \\
\hline
$0.94$  & 2.08 & 1.85 & 1.01 & 0.31 & 0.17 & 1.02 & 0.33 \\
\hline
\hline
\end{tabular}
\end{center}
\end{table*}

In this paper we first do the $3\times 3$ mixing with all three 
inputs, $f_2(1270)$, $f_2'(1525)$, and $f_2(1950)$, with 
(\ref{2++}). In this case we can fix all five mixing parameters, 
including the chromon mass $\mu$, completely. To find 
the solution, however, we have to vary the masses a 
bit. Changing the masses of $f_2(1270)$, $f_2'(1525)$, 
and $f_2(1950)$ to 1275, 1515, and 1944 MeVs, we obtain 
Table \ref{t2+1} which suggests the chromon mass to be 
around 920 MeV. 

One might worry that $\mu\simeq$ 920 MeV of Table \ref{t2+1} is a bit too large. But remember that here 
the $2^{++}$ chromoball is $^5S_2$ state in which 
the spin of two chromons are parallel. And the spin-spin interaction could have made the chromon mass large. 
So the large chromon mass here actually could be interpreted to include the energy coming from 
the spin-spin interaction. 

The result tells that $f_2(1270)$ is predominantly 
the $u\bu+d\bd$ state, $f_2'(1525)$ is predominantly 
the $s\bs$ state, and $f_2(1950)$ is predominantly 
the chromoball state. This agrees well with the result of 
the preceding paper, and is consistent with the PDG 
view \cite{pdg,prd15}. 

Notice that the table gives us 
$R(f_2'(1525)/f_2(1270))\simeq$ 0.41 which agrees 
well with the PDG value 0.36, but  
$R(f_2(1950)/f_2(1270))$ becomes 2.33 which is a little 
larger than the PDG value (\ref{pdg2}). But we find that 
we could reduce this number by changing the mass of 
$f_2(1525)$ to around 1490 MeV. With this change of 
the input, the chromon mass is reduced to around 840 MeV 
and $R(f_2(1950)/f_2(1270))$ becomes around 0.57. 

Now, remember that here we have $f_0(2010)$ just above 
2 GeV, and it would be unfair to exclude this in the mixing.
So we consider the $4\times 4$ mixing with the four mass 
eigenstates and (\ref{2++}) as the input, and obtain 
Table \ref{t2+2}. Here again we have changed the input 
masses a little, to 1275, 1500, 1944, and 2100 MeVs, 
to find the solution.  

Remarkably the result in Table \ref{t2+2} is very similar 
to the Table \ref{t2+1}. Although the numbers are different,
the general feature is the same. Here again $f_2(1270)$ 
becomes predominantly the $u\bu+d\bd$ state, 
$f_2'(1525)$ becomes predominantly the $s\bs$ state, 
and $f_2(1950)$ becomes predominantly a chromoball 
state. The only new thing is that $f_2(2010)$ becomes 
the second chromoball state,  so that we can interpret
$f_2(1950)$ and $f_2(2010)$ to be predominantly 
the $^5S_2$ and $^1D_2$ chromoballs. 

The main difference between the two tables is the $J/\psi$ 
relative radiative decay ratio. This is because the ratio
is very sensitive to the chromoball contents of the physical
states, so that a small change of the chromoball contents
influence the ratio significantly. In Table \ref{t2+2} 
the radiative decay ratios turn out to be larger then 
the PDG values (\ref{pdg2}). But we find that the ratios 
could be reduced to PDG values by changing the mass of 
$f_2'(1525)$ to around 1490 MeV. 

We can do the $4\times 4$ mixing with the three mass 
eigenstates below 2 GeV and (\ref{2++}) as the input, 
and try to predict the fourth state. The result is shown 
in Table \ref{t2+3}. But here again we have changed 
the mass eigenvalues to 1275, 1500, and 1944 MeVs
to obtain the solutions. 

Remarkably it predicts that the mass of the fourth state 
is around 2100 MeV, which we can identify to be 
$f_2(2010)$. With this identification Table \ref{t2+3} becomes very similar to Table \ref{t2+2}, which confirms that $f_2(1270)$ is predominantly the $u\bu+d\bd$ state, $f_2'(1525)$ is predominantly the $s\bs$ state, but $f_2(1950)$ and $f_0(2010)$ are predominantly 
the $^5S_2$ and $^1D_2$ chromoballs. 

\begin{table*}[htbp]
\caption{\label{t0-1} The numerical analysis of 
the $5\times5$ mixing in the $0^{-+}$ channel. 
Here we have used $\eta'(958)$, $\eta(1275)$, 
$\eta(1405)$, $\eta(1475)$, and 
$R(\eta(1475)/\eta'(958))=0.95$ as the input. 
The fifth state could be interpreted as $\eta(548)$.}
\begin{center}
\begin{tabular}{cccccccccccccccccccccccccccccc}
\hline
\hline
$\mu$ & $m_5$ & \multicolumn{5}{c} {$m_1=\eta'(958)$}
& \multicolumn{5}{c} {$m_2=\eta(1295)$} 
& \multicolumn{5}{c} {$m_3=\eta(1405)$} 
& \multicolumn{5}{c} {$m_4=\eta(1475)$}\\
$~$ &~& $u+d$ & $s$ & $G_1$ & $G_2$ & $G_3$ & $u+d$ 
& $s$ & $G_1$ & $G_2$ & $G_3$ & $u+d$ & $s$ & $G_1$
& $G_2$ & $G_3$ & $u+d$ & $s$ & $G_1$ & $G_2$ & $G_3$\\
\hline
$0.58$ & 0.52 & 0.18 & 0.37 & 0.43 & 0.01 & 0.01 & 0.19 
& 0.18 & 0.53 & 0.06 & 0.05 & 0.02 & 0.01 & 0.01 & 0.92 
& 0.04 & 0.04 & 0.03 & 0.02 & 0.02 & 0.90\\
\hline
$0.58$ & 0.52 & 0.18 & 0.36 & 0.44 & 0.01 & 0.00 & 0.17 
& 0.16 & 0.50 & 0.15 & 0.02 & 0.05 & 0.04 & 0.04 & 0.82 
& 0.05 & 0.02 & 0.02 & 0.01 & 0.02 & 0.93\\
\hline
$0.58$ & 0.51 & 0.18 & 0.37 & 0.43 & 0.00 & 0.01 & 0.19 
& 0.18 & 0.54 & 0.00 & 0.08 & 0.00 & 0.00 & 0.00 & 1.00 
& 0.00 & 0.04 & 0.03 & 0.02 & 0.00 & 0.91\\
\hline
$0.58$ & 0.52 & 0.17 & 0.35 & 0.46 & 0.02 & 0.00 & 0.14 
& 0.13 & 0.44 & 0.30 & 0.00 & 0.12 & 0.10 & 0.10 & 0.68 
& 0.00 & 0.00 & 0.00 & 0.00 & 0.00 & 1.00\\
\hline
\hline
\end{tabular}

\begin{tabular}{ccccccccccccccccccc}
\hline
\hline
$\mu$ & $m_5$ &\multicolumn{5}{c} {$m_5$} & $R(m_2/m_1)$ 
& $R(m_3/m_1)$ & $R(m_4/m_1)$ & $R(m_5/m_1)$ & A & $\nu$ 
& $\nu'$ & $\delta$& $\delta'$ & $\mu_0$ \\
~&~& $u+d$ & $s$ & $G_1$& $G_2$ & $G_3$ &~&~&~&~
&~&~&~&~&~&~&~&\\
\hline
$0.58$ & 0.52 & 0.58 & 0.41 & 0.01& 0.00 & 0.00 & 0.90 & 1.12 
& 0.95 & 0.03 & 0.36 & 0.40 & -0.17 & 0.62 & -0.89 & 0.487 \\
\hline
$0.58$ & 0.52 & 0.58 & 0.41 & 0.01 & 0.00 & 0.00 & 0.93 & 1.02 
& 0.95 & 0.03 & 0.37 & 0.41 & -0.17 & 0.58 & -0.87 &0.489 \\
\hline
$0.58$ & 0.51 & 0.58 & 0.41 & 0.01 & 0.00 & 0.00 & 0.89 & 1.17 
& 0.95 & 0.03 & 0.36 & 0.40 & 0.01 & 0.63 & -0.91 & 0.485 \\
\hline
$0.58$ & 0.52 & 0.57 & 0.42 & 0.01 & 0.01 & 0.00 & 0.97 & 0.84 
& 0.95 & 0.03 & 0.38 & 0.43 & -0.01 & 0.52 & -0.85 & 0.492 \\
\hline
\hline
\end{tabular}
\end{center}
\end{table*}

So, all in all the mixing in the $2^{++}$ channel seems to 
work fine, and the upshot of our mixing is that $f_2(1950)$ 
and $f_0(2010)$ are predominantly the chromoball states. 
On the other hand, it is good to remember that there are
different suggestions in the literature. Clearly there have been claims that $f_2(1270)$ and $f_2'(1525)$ are the $q\bq$ states 
as PDG suggests \cite{f212701,f212702,f212703}. On the other hand, there have been assertions that they can be viewed 
as molecular states \cite{f212704,f212705,f212706,f212707}. 
So we need more time to understand the physical contents 
of the $2^{++}$ states clearly.

But what really makes the mixing analysis complicated is 
the fact that experimentally we have five unestablished 
states here, $f_2(1430)$, $f_2(1565)$, $f_2(1640)$, 
$f_2(1810)$, and $f_2(1910)$ \cite{pdg}. Some of them 
could turn out to be real states and make the mixing 
unreliable. So we need more experimental clarification 
on the unestablished states. Even if all remain unestablished, 
we have to explain why there are so many unestablished 
states in this channel. 

\subsection{$0^{-+}$ channel}

This channel has attracted special attention because of
the octet-singlet mixing, the U(1) problem, PCAC etc.
In this channel we have five established states below 
2 GeV, $\eta(548)$, $\eta'(958)$, $\eta(1295)$, 
$\eta(1405)$, and $\eta(1475)$, and one unestablished state $\eta(1760)$. 

In the preceding paper we discussed the $4\times 4$ mixing
of two chromoball states and two $q\bq$ states, using 
\bea
&E=m^2(\pi),~~~\pi=\pi(140), \nn \\
&\Delta =2(m^2(K)-m^2(\pi)),~~~~K=K(498),
\label{0-+}
\eea
with $\eta'(958)$, $\eta(1405)$, and $\eta(1760)$ as 
the input. 

The result showed that the mass of the fourth 
physical state becomes around 550 MeV, which could 
be interpreted to be $\eta(548)$. In this case $\eta(548)$ 
turns out to be a mixture of $u\bu+d\bd$ and $s\bs$, 
while $\eta'(958)$ becomes predominantly a $gg$ 
chromoball \cite{prd15}. 

This is not satisfactory and not in line with PDG, which 
interprets $\eta'(958)$ as predominantly a $q\bq$ state. 
There are other problems. For example, the physical 
contents of $\eta(1405)$ and $\eta(1760)$ depended 
very much on the chromon mass. 

Moreover, the $J/\psi$ radiative decay ratios in this 
table do not agree well with PDG values. Indeed, 
experimentally PDG has a new data 
\bea
&J/\psi\rightarrow\gamma\eta(548)
\simeq (1.104\pm 0.034) \times 10^{-3},  \nn\\
&J/\psi\rightarrow\gamma\eta'(958)
\simeq (5.15\pm 0.16) \times 10^{-3},  \nn\\
&J/\psi\rightarrow\gamma\eta(1405/1475)
\simeq 4.9\times 10^{-3},  
\eea
which tells 
\bea
&R(\eta'(958)/\eta(548))\simeq 4.66,\nn\\
&R(\eta(1405/1475)/\eta(548))\simeq 4.44.
\label{0-rd}
\eea
So we need to explain this.

But a most critical defect of the $4\times 4$ mixing 
is that it can not explain all five physical states. 
This is the critical shortcoming of the $4\times 4$
mixing. For this reason we discuss the $5\times 5$ 
mixing in the following which could explain all five 
established states. 

\begin{table*}[htbp]
\caption{\label{t0-2} The numerical analysis of 
the $3\times 3$ mixing in the $0^{-+}$ channel. Here we choose $\eta'(958)$ and $\eta(1405)$ as the input 
and vary the mass of $\eta(548)$to obtain the table. 
No solution can be found when $m(\eta(548))>$ 
541 MeV.}
\begin{center}
\begin{tabular}{cccccccccccccccc}
\hline
\hline
 $m(\eta(548))$&$A$&$\nu$&$\mu$ 
& \multicolumn{3}{c} {$m_1=\eta(548)$} 
& \multicolumn{3}{c} {$m_2=\eta'(958)$} 
& \multicolumn{3}{c} {$m_3=\eta(1405)$} 
& $R(m_3/m_2)$ & $R(m_1/m_2)$\\
$~$ &~&~&~& $u+d$ & $s$ & $G$ & $u+d$ & $s$ & $G$ 
& $u+d$ & $s$ & $G$ &~&~\\
\hline
$510$ & 0.34 & 0.51 & 0.64 & 0.60 & 0.39 & 0.01 & 0.23 
& 0.47 & 0.30 & 0.17 & 0.14 & 0.69 & 1.2 & 0.05\\
\hline
$520$ & 0.41 & 0.55 & 0.60 & 0.56 & 0.43 & 0.01 & 0.16 
& 0.33 & 0.51 & 0.28 & 0.24 & 0.48 & 0.49 & 0.03\\
\hline
$530$ & 0.50 & 0.48 & 0.55 & 0.51 & 0.48 & 0.01 & 0.09 
& 0.19 & 0.72 & 0.40 & 0.34 & 0.27 & 0.19 & 0.02\\
\hline
$540$ & 0.58 & 0.22 & 0.49 & 0.47 & 0.53 & 0.00 & 0.01 
& 0.03 & 0.95 & 0.52 & 0.43 & 0.04 & 0.02 & 0.003\\
\hline
\hline
\end{tabular}
\end{center}
\end{table*}

Consider the following mixing matrix  
\bea
M^2=\left( \begin{array}{ccccc}
E+\dfrac{2}{3}\Delta  & -\dfrac{\sqrt{2}}{3} \Delta 
& 0 & 0 & 0 \\
-\dfrac{\sqrt{2}}{3}\Delta & E+\dfrac{1}{3}\Delta 
+3A & \nu_1 & \nu_2 & \nu_3 \\
0  &  \nu_1 & G_1 & 0 & 0 \\
0  & \nu_2 & 0 & G_2 & 0 \\
0  & \nu_3 & 0 & 0 & G_3 \end{array} \right),
\label{massm5}
\eea
which describes the mixing of three chromoball states
with two quarkoniums below 2 GeV. This has nine 
parameters. Now, normally we could choose seven 
inputs, (\ref{0-+}) and five mass eigenvalue, and treat 
$\mu$ as a free parameter. In this case we need one 
more constraint, and might impose the following 
constraint
\bea
&\nu_1=\nu,~~~\nu_2=\dfrac{\nu_1+\nu_3}{2},
~~~\nu_3=\nu',
\label{c10-+}
\eea 
just for simplicity. 

But here we choose a slightly different input. We 
choose four mass eigenstates, $\eta'(958)$, 
$\eta(1295)$, $\eta(1405)$, $\eta(1475)$, and 
$R(\eta(1475)/\eta'(958))=0.95$ of (\ref{0-rd}) 
in stead of $\eta(548)$. With this we could predict 
the mass of the fifth physical state. The reason is 
that, as we have pointed out the mathematical 
equations which we need to solve to diagonalize 
the mass matrix are very rigid, so that we could 
not find the solution when we use the five mass 
eigenstates as the inputs. 

Assuming that $G_3$ is the $ggg$ chromoball we may 
let
\bea
&G_1=4\mu^2,~~~G_2=4\mu^2 +\delta,
~~~G_3=9\mu^2+\delta'.
\label{c20-+}
\eea
Actually this is also artificial, because here $G_1,~G_2,~G_3$
are supposed to be the mass eigenstates of three chromoballs. 
Nevertheless we adopt (\ref{c20-+}) here, because this 
could provide some insight on the chromon mass and 
their binding. 

With (\ref{c10-+}) and (\ref{c20-+}) we obtain Table \ref{t0-1}. The result shows that the mass of the fifth physical state is around 520 MeV, which could be 
identified as $\eta(548)$. In this case $\eta(548)$ 
turns out to be a mixture of $u\bu+d\bd$ and $s\bs$, 
while $\eta'(958)$ becomes largely a mixture of 
$s\bs$ and a $gg$ chromoball, with less than 20\% contamination of $u\bu+d\bd$. And $\eta(1295)$ is 
made of more than 50\% $gg$ chromoball and less 
than 20\%  $u\bu+d\bd$ and $s\bs$ each. 

But remarkably, the table shows that $\eta(1405)$ 
and $\eta(1475)$ are mainly the chromoball states.
Moreover, the $J/\psi$ radiative decay ratios 
$R(\eta(1405)/\eta'(958))$ is perfect, although 
$R(\eta'(958)/\eta(548))$ looks a bit larger. 
This looks interesting and reasonable, considering 
the fact that we have imposed the {\it ad hoc} 
constraints (\ref{c10-+}) and (\ref{c20-+}).

Notice that $\delta'$ turns out to be negative, which shows 
that the mass of the three chromon bound state is smaller 
than the sum of the chromon masses. This might be understood to imply that the three chromon binding is 
quite strong. On the other hand this could also be an artefact of (\ref{c20-+}). For instance we could introduce 
a new chromon mass $\mu_0$ with $G_3=3\mu_0^2$ as shown in the table, and avoid the negative binding energy.  
  
In the literature, of course, we have different views. 
The popular view that PDG endorses is that $\eta(548)$ 
and $\eta'(958)$ are predominantly the $u\bu+d\bd$ 
and $s\bs$ states, and that $\eta(1295)$ and $\eta(1475)$ 
are the first radial excitations of $\eta(548)$ and 
$\eta'(958)$ \cite{e14751,e14752,e14753}. But it is widely 
agreed that $\eta(1405)$ is indeed a pseudo-scalar 
glueball \cite{e14051,e14052,e14053,e14054,e14055,e14056}. 
This is endorsed by PDG and by our analysis, although 
there exists a lattice result which might contradict with 
this view \cite{lqcd2}. 

Our result implies that the spectrum of the light 
pseudo-scalar mesons could be understood within 
the context of the quarkonium-chromoball mixing. 
Nevertheless, the idea that $\eta(1295)$ and 
$\eta(1475)$ could be the radial excitations of 
$\eta(548)$ and $\eta'(958)$ should be taken 
seriously \cite{e14751,e14752,e14753}. 

To see how this popular view fares in our quark and 
chromon model, we consider the $3\times 3$ mixing 
with only three physical states, $\eta(548)$, $\eta'(958)$, 
and $\eta(1405)$, excluding the supposedly radially 
excited states $\eta(1295)$ and $\eta(1475)$. Normally 
in the $3\times 3$ mixing we could use the three masses
and (\ref{0-+}) as the input to diagonalize the mass matrix, 
but in this case we could not find the solution. So we choose 
only two mass eigenvalues, $\eta'(958)$ and $\eta(1405)$, 
and vary the mass of $\eta(548)$. With this we obtain 
Table \ref{t0-2}.

Interestingly, when the mass of $\eta(548)$ becomes 
510 MeV, the radiative decay ratio $R(\eta(1405)/\eta(958))
\simeq$ 1.2 becomes close to the experimental value 0.95. 
In this case $\eta(548)$  becomes 60\% $u\bu+d\bd$
and 39\% $s\bs$, but $\eta'(958)$ becomes a mixture 
of 47\% $s\bs$ and 30\% $gg$. And $\eta(1405)$ 
becomes predominantly (69\%) a chromoball. 

To understand the physical meaning of Table \ref{t0-2}, 
we notice that the physical contents depends very much 
on the mass of $\eta(548)$. Moreover, as the mass 
approaches to the physical value 548 MeV, $\eta'(958)$ 
becomes predominantly the glueball. 

This is troublesome, and does not seem to support 
the PDG view (that $\eta(1295)$ and $\eta(1475)$ are 
the radial excitations of $\eta(548)$ and $\eta'(958)$) 
at all. This implies that our result shown in 
Table \ref{t0-1} is at least as good as the PDG view, 
although this matter has to be studied more carefully. 
    
In this section we have extended and improved 
the numerical analysis of the quarkonium-chromoball 
mixing of the preceding paper in three channels 
$0^{++}$, $2^{++}$, and $0^{-+}$ below 2 GeV, 
based on our quark and chromon model. Although 
the numerical results are still inconclusive, the results 
in this paper seem to work better. 

Theoretically it must be clear that the numerical 
mixing should be regarded as an approximation. 
Moreover, technically the equation we need to 
solve to diagonalize the mass matrix is very rigid 
and sensitive to the {\it ad hoc} constraints we 
have imposed.

With these shortcomings it is natural that our results 
are not perfect. Nevertheless, it is fair to say that 
the above mixing analysis does show that the quark 
and chromon model provides a conceptually simple 
way to understand the glueballs and their mixing with 
quarkoniums. 

\section{Discussions}

One of the main problems in hadron spectroscopy 
has been the identification of the glueballs. In this 
paper we have made the numerical analysis of 
chromoball-quarkonium mixing to identify the glueballs, based on the quark and chromon model obtained by 
the Abelian decomposition \cite{prd15}. Our mixing 
analysis is a rough approximation, but it does confirm 
that the glueballs (i.e., the chromoballs) play 
a fundamental role in the hadron spectroscopy, 
although in general (except for the oddballs) they 
exist as mixed states. In fact the analysis tells that 
it is simply impossible to understand the meson 
spectroscopy without them.

Our analysis was able to pinpoint the glueball candidates
below 2 GeV successfully. Indeed our result strongly 
indicates that $f_0(1500)$ in the $0^{++}$ sector, $f_2(1950)$ in the $2^{++}$ sector, and $\eta(1405)$ 
and $\eta(1475)$ in the $0^{-+}$ sector become predominantly the glueballs. Some of them have been suggested to be the glueball states before, but some of them (e.g., $\eta(1475)$) are our suggestion.

In our mixing analysis we have also tried to settle other 
unresolved issues. First, in the $0^{++}$ sector 
an important issue is what is the isotriplet $q\bq$ 
partner of the isosinglet $q\bq$. There are two 
contending views. The popular view endorsed by 
PDG is that $a_0(1450)$ is the isotriplet partner, but 
the opposite view suggests that $a_0(980)$ is 
the isotriplet parner \cite{pdg,prd15}. The popular 
view appears intuitively strange because, if this is so, 
the strange flavour octet partner $K_0^*(1430)$ 
becomes lighter than $a_0(1450)$. So it is important 
to find out which view is correct, and why. We tried 
to resolve this issue in our mixing. Unfortunately our 
analysis could not provide a conclusive answer on this, 
but it does imply that the opposite view is not completely excluded yet.    

Another issue in this sector is the nature of $f_0(500)$,
which has been a big mystery \cite{f5001,f5002,f5003,
f5004,f5005,f5006,f5007,f5008,f5009,f50010,f50011}. 
In the quark and chromon model, the chromons are supposed to be the consituent gluons, but logically 
we can not exclude the possibility that the neurons 
could also form a loosely bound state. In this paper 
we discussed this possibility. Our analysis suggests 
that $f_0(500)$ could be viewed a neuroball state. 
And this is independent of which state we choose to be 
the isotriplet partner. In our quark and chromon model 
the neuroballs (if exist) should look very much like 
loosely bound states of two (or three) $q\bq$ mesons 
or $gg$ chromoballs, and $f_0(500)$ nicely fits in this picture. Remarkably, this is consistent with the popular 
view advocated by many authors \cite{f5001,f5002,
f5003,f5004,f5005,f5006,f5007}. But we emphasize 
that in detail two views are different. The one interprets $f_0(500)$ to be a glueball, but the other interprets it a molecular state.    

A related issue is whether the monopole condensation 
in QCD could generate a $0^{++}$ vacuum fluctuation 
mode or not \cite{prd80,prl81,prd15}. Theoretically 
this, of course, is a fundamental question. If the answer
turns out to be in the affirmative, $f_0(500)$ would 
be a natural candidate of the vacuum fluctuation. 
This is a very interesting and attractive possibility which 
warrants further study. Here we just emphasize that our analysis does not exclude this possibility. 

The mixing in the $2^{++}$ sector is rather straightforward 
because there are no controversial issues here. Here 
we have three physical states below 2 GeV, and our 
result tells that $f_2(1275)$ and $f'_2(1525)$ are 
the $u\bu+d\bd$ and $s\bs$ states, respectively. This, 
of course, is in line with the PDG interpretation \cite{pdg}. 
Moreover, our result tells that $f_2(1950)$ is predominantly 
the chromoball state, which agrees with our result in 
the preceding paper \cite{prd15}. 

This sounds all very nice, but we have to swallow 
this with a grain of salt. The problem is that in this 
sector PDG shows that there are five unestablished 
states, and some of them could turn out to be real. 
And it is quite possible that this could give us 
a serious impact on the mixing analysis. 

Finally, in the $0^{-+}$ sector an important issue is 
whether $\eta(1295)$ and $\eta(1750)$ are the radial 
excitations of $\eta(548 )$ and $\eta'(958)$ or not 
\cite{e14751,e14752,e14753}. Our mixing analysis 
provides a different picture. Our result tells that 
$\eta(1295)$ can be viewed as a mixed state made of 
more than 50\% $gg$ chromoball and less than 20\%  
$u\bu+d\bd$ and $s\bs$ each, and $\eta(1475)$ is 
mainly the $ggg$ chromoball state.  This looks very 
interesting and reasonable, although we have yet 
to see which view is correct. 

One of the problems in the mixing analysis is that 
the mathematical equations to diagonalize the mass 
matrix are very rigid and sensitive to the input data.
This is troublesome because in reality we often do not 
have enough input data. This has forced us to impose 
{\it ad hoc} constraints like (\ref{c0+55}) and (\ref{c10-+}) 
which may have distorted the reality. But this is a technical 
problem we could avoid when enough experimental data 
become available.

Independent of the details, however, we emphasize 
the conceptual simplicity and clarity of the quark and 
chromon model. As a natural generalization of the quark 
model it tells what are the glueballs made of and how they 
mix with quarkoniums without ambiguity. Most importantly, 
it provides us the general framework of the hadron 
spectroscopy in simple and clear terms. 

Of course there are other models of glueball, in particular 
the constituent gluon model, which allow similar mixing 
analysis. In fact, superficially our mixing analysis is almost 
identical to the mixing in this model. As we have emphasized, however, the constituent model has 
the critical defect that it can not tell exactly what are 
the constituent gluons. In comparison our model 
tells what are the constituent gluons and what are 
the binding gluons which bind the constituent gluons. 
This is because our model is based on different logic, 
that QCD is made of two types of gluons which play 
different roles. No other model is based on this 
fact. 

To amplify this point consider the so-called ``model 
independent" calculations of gluball spectrum, the QCD 
sum rule approach \cite{shif} and the lattice calculation 
\cite{lqcd1,lqcd2}. It has been assumed that these 
calculations are based on ``the first principles" of QCD 
and thus regarded as model independent. But we have 
to know what is the first principles of QCD before we 
know how to calculate the glueball spectrum. As we 
have emphasized, the Abelian decomposition reveals 
the hidden principles of QCD, which makes 
the old-fashioned first principles obsolete.  

For instance, in the QCD sum rule approach people have
been calculating the glueball mass with the conventional 
current operators made of two gauge field strengths, claiming that this is based on the model independent 
first principles. However, the Abelian decomposition 
tells that actually there is a new and much simpler way 
to calculate the glueball mass, with the gauge invariant current operators made of two chromons. And 
obviously the two methods will give us different 
results. 

Exactly the same way, in the lattice calculation we 
can construct the glueballs implementing the Abelian 
decomposition on lattice or without implementing it. 
And again we get different results \cite{cundy1,cundy2}. 
Clearly in the conventional lattice glueball calculations,
the ingredient of the glueballs is two or three gauge 
field strengths. In comparison, in our approach 
the ingredient of the glueballs is two or three chromons,
and obviously the chromons are totally different from 
the gauge field strengths. Consequently the two 
calculations should have different results. 

These two examples clearly tells that we must understand 
the first principles of QCD first, before we actually make 
the ``model independent" calculations. As we have explained in the first part of the paper, the Abelian decomposition allows us to do that. And this is not 
a conjecture, but mathematically a well established 
fact in QCD \cite{prd80,prl81,prd81,duan,kondo1,kondo2,
cundy1,cundy2,prd01a,prd13,ijmpa14,fadd,shab,gies,zucc}. This is the advantage of the quark and chromon model.  

Before we close we emphasize that the quark and chromon 
model is not just a theoretical proposal. The underlying 
proposition of the model that there exist two types of gluons  
could be tested directly by experiment. We already have 
enough knowledge on how to differentiate the gluon jet from 
the quark jet experimentally \cite{jet1,jet2,jet3,jet4,jet5}. 
Moreover, there has been a new proposal on how to separate different types of jets at LHC \cite{jet6}. Using these knowledge we could actually confirm the existence 
of two types of gluon jets experimentally. So we do have 
an unmistakable way to justify the quark and chromon 
model experimentally.

Obviously our numerical results in this paper are not 
perfect, and can not explain everything. Nevertheless 
they do demonstrate that the quark and chromon model 
is at least as good as any other model in the literature 
which describes the glueballs and their mixing with 
quarkoniums. Moreover, the numerical mixing analysis 
is not the only the application of our model. The next 
application would be to implement the Abelian decomposition to the QCD sum rule and the lattice QCD calculations, and obtain a better understanding of 
glueballs. The work in these directions are in progress.

{\bf ACKNOWLEDGEMENT}

~~~The work is supported in part by National Natural Science Foundation of China (Grant No. 11575254), 
China Scholarship Council, and Basic Science Research Program through the National Research Foundation 
of Korea funded by the Ministry of Education (Grants 
2015-R1D1A1A0-1057578 and 2018-R1D1A1B0-7045163), 
and by the Center for Quantum Spacetime at Sogang University.


\begin{thebibliography}{99}
\bibitem{lambda} Our notation of the $\lambda$-matrices 
is the same as the Gellman's notation, except that 
$\lambda_5$ has the opposite signature.
\end{thebibliography}
\end{document}